\documentclass[a4paper,12pt]{article}
\pdfoutput=1
\usepackage{jheppub}
\usepackage{amsmath,amsfonts,graphicx,yfonts,mathabx}
\usepackage{amssymb}
\usepackage{epsfig, graphicx, yfonts, float}
\usepackage{tikz}
\usetikzlibrary{decorations.pathmorphing}
\usepackage{capt-of}
\usepackage{caption}

\newcommand{\be}{\begin{equation}}
\newcommand{\ee}{\end{equation}}
\newcommand{\bea}{\begin{eqnarray}}
\newcommand{\eea}{\end{eqnarray}}

\newcommand{\mt}[1]{\textrm{\tiny #1}}

\def\rh {r_\mt{H}}

\subheader{\begin{flushright}
UTTG-15-18
\end{flushright}}

\title{Chaos and entanglement spreading in a non-commutative gauge theory}

\author{Willy Fischler$^{\dag}$,}
\affiliation{$^{\dag}$Theory Group, Department of Physics and Texas Cosmology Center, The University of Texas at Austin,
Austin, TX 78712, USA}
\author{Viktor Jahnke$^{\ddag}$}
\affiliation{$^{\ddag}$Departamento de F\'isica de Altas Energias, Instituto de Ciencias Nucleares, Universidad Nacional Aut\'onoma de M\'exico, Apartado Postal 70-543, CDMX 04510, M\'exico}
\author{and Juan F. Pedraza$^{\star}$}
\affiliation{$^{\star}$Institute for Theoretical Physics, University of Amsterdam, Science Park 904, 1098 XH Amsterdam, Netherlands}
\emailAdd{fischler@physics.utexas.edu}
\emailAdd{viktor.jahnke@correo.nucleares.unam.mx}
\emailAdd{jpedraza@uva.nl}

\abstract{Holographic theories with classical gravity duals are maximally chaotic: they saturate a set of bounds on the spread of quantum information. In this paper we question whether non-locality can affect such bounds. Specifically, we consider the gravity dual of a prototypical theory with non-local interactions, namely, $\mathcal{N}=4$ non-commutative super Yang Mills. We construct shock waves geometries that correspond to perturbations of the thermofield double state with definite momentum and study several chaos related properties of the theory, including the butterfly velocity, the entanglement velocity, the scrambling time and the maximal Lyapunov exponent. The latter two are unaffected by the non-commutative parameter $\theta$, however, both the butterfly and entanglement velocities increase with the strength of the non-commutativity. This implies that non-local interactions can enhance the effective light-cone for the transfer of quantum information, eluding previously conjectured bounds encountered
in the context of local quantum field theory. We comment on a possible limitation on the retrieval of quantum information imposed by non-locality.}

\begin{document}

\maketitle

\flushbottom

\section{Introduction} \label{sec-intro}

\subsection{Probes of quantum chaos}

Recent studies of many-body quantum chaos have shed light into the inner-working mechanisms of the gauge/gravity duality \cite{duality1,duality2,duality3}. For example, the characteristic velocity of the butterfly effect is known to play an important role in determining the bulk causal structure \cite{Qi-2017}, while the saturation of the maximal Lyapunov exponent might be a necessary condition for the existence of a gravity dual \cite{bound-chaos,Kitaev2014}. There are also some interesting proposals connecting chaos and hydrodynamics \cite{Blake1,Blake2,Davison:2016ngz,Blake:2017qgd,Grozdanov:2017ajz,Blake-2017,Grozdanov:2018atb}, and chaos and the spread of quantum entanglement \cite{Mezei-2016,Mezei-2016v2}.

One way to diagnose chaos in quantum many-body systems is to consider the influence of an early perturbation $V$ on the later measurement of some other operator $W$. Such an effect is encoded in the quantity \cite{larkin}
\begin{equation}\label{commutator}
C(t)=- \langle [W(t),V(0)]^2 \rangle\,,
\end{equation}
where $\langle \cdots \rangle=Z^{-1}\text{tr}[e^{\beta H}\cdots]$ denotes the thermal expectation value at temperature $T=\beta^{-1}$.
For chaotic systems the expected behavior is the following \cite{BHchaos1,BHchaos2}
\[ C(t) \sim
  \begin{cases}
    N^{-2}      & \quad \text{for }\,\, t < t_d\,,\\
    N^{-2} \exp \left( \lambda_L t\right) & \quad \text{for}\,\, t_d <\!\!< t <\!\!< t_*\,,\label{chaosresult}\\
    \mathcal{O}(1) & \quad \text{for}\,\, t > t_*\,,
  \end{cases}
\]
where $N^2$ is the number of degrees of freedom of the system. Here, we have assumed $V$ and $W$ to be few-body Hermitian operators normalized such that $\langle V V\rangle=\langle W W\rangle=1$. The exponential growth of $C(t)$ is characterized by the Lyapunov exponent $\lambda_L$ and takes place at intermediate time scales bounded by the dissipation time $t_d$ and the scrambling time $t_*$. The dissipation time characterizes the exponential decay of two-point correlators, e.g., $\langle V(0) V(t) \rangle \sim e^{-t/t_d}$, while the scrambling time $t_* \sim \lambda_L^{-1} \log N^2$ is defined as the time at which $C(t)$ becomes of order $\mathcal{O}(1)$ \cite{scrambling1, scrambling2}. The behavior of $C(t)$ can be understood in terms of the expansion of $V$ in the space of degrees of freedom. Under time evolution, the operator $V$ gets scrambled with an increasing number of degrees of freedom and this causes $C(t)$ to grow. Eventually, $V$ gets scrambled with \emph{all} degrees of freedom available in the system and, as consequence, $C(t)$ saturates to a constant $\mathcal{O}(1)$ value.

In holographic theories, the dissipation time is controlled by the black hole quasinormal modes, so one generally expects $t_d \sim \beta$ for low dimension operators. On the other hand, the scrambling time for black holes is found to be $t_* \sim \beta \log N^2$. For general quantum systems with such a large hierarchy between these two time scales, $t_d <\!\!< t_*$, the Lyapunov exponent was shown to have a sharp upper bound \cite{bound-chaos}
\be\label{lyapbound}
\lambda_L \leq \frac{2 \pi}{\beta}\,.
\ee
Interestingly, this bound is saturated by black holes in Einstein gravity, leading to the speculation that
any large $N$ system that saturates this bound will necessarily have an Einstein gravity dual, at least in the near horizon region \cite{bound-chaos,Kitaev2014}.
Such a claim triggered an enormous interest in the community, and lead to many works attempting to use the saturation of the bound as a criterion to discriminate between CFTs with potential Einstein gravity duals \cite{Roberts:2014ifa,Fitzpatrick:2016thx,Michel:2016kwn,Caputa:2016tgt,Perlmutter:2016pkf,Turiaci:2016cvo,Caputa:2017rkm}. However, it was recently proved that this criterion by itself is  insufficient (albeit necessary) to guarantee a dual description with gravitational degrees of freedom \cite{deBoer:2017xdk}.

A further diagnose of quantum chaos comes from considering the response of the system to arbitrary local perturbations. This effect can be studied by upgrading the commutator in (\ref{commutator}) to
\be
C(t,\vec{x})=- \langle [W(t,\vec{x}),V(0)]^2 \rangle\,.
\label{eq-C(t,x)}
\ee
Calculations for holographic systems \cite{BHchaos3,BHchaos4} and the SYK chain suggested that for chaotic systems, the exponential growth regime in (\ref{chaosresult}) generalizes to:
\begin{equation}
C(t,\vec{x}) \sim N^{-2} \exp \left[ \lambda_L\left(t-\frac{|\vec{x}|}{v_B}\right) \right]\,, \,\,\,\,\text{for}\,\, |\vec{x}|>\!\!>1\,.
\label{eq-C(t,x)}
\end{equation}
The butterfly velocity $v_B$ characterizes the rate of expansion of the operator $V$ in space. This quantity defines an emergent light cone, defined by $t-t_* = |\vec{x}|/v_B$. Within the cone, i.e. for $t-t_* >|\vec{x}|/v_B$, one has that $C(t,\vec{x}) \sim \mathcal{O}(1)$, whereas outside the cone, for $t-t_* <|\vec{x}|/v_B$, one has $C(t,\vec{x}) \approx 0$. Interestingly, in \cite{Roberts:2016wdl} it was argued that $v_B$ acts as a low-energy Lieb-Robinson velocity, which sets a bound for the rate of transfer of quantum information. In \cite{Mezei-2016v2} it was proved that for asymptotically AdS black holes in two-derivative (Einstein) gravity, satisfying null energy condition (NEC), the butterfly velocity is bounded by
\begin{equation}\label{boundvB}
v_B \leq v_B^\mt{Sch}=\sqrt{\frac{d}{2(d-1)}}\,,
\end{equation}
where $v_B^\mt{Sch}$ is the value of the butterfly velocity for a $(d+1)$-dimensional AdS-Schwarzschild black brane. It is tempting to conjecture that (\ref{boundvB}) might be a bound for any (local) QFT, in the same sense as the bound for the Lyapunov exponent (\ref{lyapbound}). However, (\ref{boundvB}) was shown to fail for higher derivative gravities \cite{BHchaos3}, as well as for anisotropic theories in Einstein gravity \cite{Giataganas:2017koz,Jahnke-2017}, which is reminiscent of the well-known violation of the shear viscosity to entropy density ratio \cite{Brigante:2007nu,Brigante:2008gz,Camanho:2010ru,Erdmenger:2010xm,Rebhan:2011vd,Jahnke:2014vwa}. In such cases, however, $v_B$ is still bounded from above and never reaches the speed of light $c=1$, provided that the theory respects causality.\footnote{The butterfly velocity can exceed the speed of light if causality is violated. For instance, Gauss-Bonnet gravity in $d=4$ dimensions has $v_B>1$ for $\lambda_{\text{GB}}<-0.75$. However, causality only holds for $\lambda_{\text{GB}}>-0.19$ \cite{Camanho:2009vw,Buchel:2009sk} (furthermore, it requires an infinite tower of extra higher spin fields \cite{Camanho:2014apa}).} Naively, one would expect the speed of light to define a region of causal influence in a relativistic system. However, as clarified in \cite{Qi-2017}, when we only have access to a subset of the Hilbert space, the propagation velocity of causal influence is generically smaller than the speed of light. So, we usually have $v_B < 1$ because the butterfly velocity characterizes the velocity of causal influence in a subset of the Hilbert space defined by the thermal ensemble, i.e. the states with a fixed energy density. Indeed, the authors of \cite{Qi-2017} showed that, for any asymptotically AdS geometry in two-derivative gravity, the butterfly velocity is bounded by the speed of light, i.e.
\be\label{boundvB2}
v_B \leq 1\,,
\ee
as it should for a theory with a (Lorentz invariant) UV fixed point.

A natural question one can ask is whether non-local interactions can lead to a violation of either the Lyapunov exponent bound (\ref{lyapbound}) or the butterfly velocity bound (\ref{boundvB2}). Since non-local interactions break Lorentz invariance, \emph{a priori} one does not expect $c$ to play a role. Furthermore, non-local theories with holographic duals, have bulk metric that are in fact non-asymptotically AdS, so the bound derived in \cite{Qi-2017} does not apply. Known examples of non-local holographic theories are, for instance, $i)$ the near  horizon  limit  of  a  stack  of D3-branes with a constant Neveu-Schwarz $B_{\mu\nu}$ \cite{Hashimoto:1999ut,Maldacena-Russo}, dual to non-commutative $\mathcal{N}=4$ super Yang Mills, $ii)$ the near  horizon  limit  of  a  stack  of D3-branes with global R-symmetry charges \cite{Bergman:2001rw}, dual to a dipole deformation of $\mathcal{N}=4$ super Yang Mills and $iii)$  the  theory  dual  to  the  near  horizon  limit  of  a  stack  of  NS5-branes \cite{Aharony:1998ub}, the so-called little string theory.

In this paper we will explore the aforementioned question in a prototypical theory with non-local interactions, namely, $\mathcal{N}=4$ non-commutative super Yang Mills.\footnote{Needless to say, in the future it would be interesting to consider other examples of non-local theories and compare with the results of this paper.}
Holography has already been useful to explore several dynamical effects of the non-locality inherent to non-commutative theories, with some surprising findings. For instance, in \cite{Edalati:2012jj} it was shown from a quasinormal mode analysis that non-commutative gauge theories display a parametrically shorter dissipation time for light probes, i.e. $t_d(\theta)<\!\!<t_d(0)$ for $T\sqrt{\theta}>\!\!>1$ and $k\sqrt{\theta}>\!\!>1$, where $\theta$ measures the strength of the non-commutativity. Heavy probes were further analyzed in \cite{Matsuo:2006ws,Fischler:2012ff} showing a qualitative reduction in the viscosity felt by the probe. Lastly, the holographic complexity was recently studied in \cite{Couch:2017yil} which, remarkably, was shown to violate the so-called Lloyd's bound at late times.

\subsection{Chaos and entanglement spreading}

One way to diagnose chaos in holographic theories is by studying the disruption of mutual information between subregions of the two boundaries in a maximally extended black brane geometry. This approach is particularly interesting because it makes a clear connection between chaos and spreading of entanglement, which is another topic that will be relevant for our discussion.

Holographically, a maximally extended (two-sided) black brane geometry is dual to a thermofield double state (TFD) of two identical copies of the theory, which we call QFT$_L$ and QFT$_R$, respectively \cite{eternalBH}. At $t=0$, the TFD state is given by
\begin{equation}
|TFD \rangle =\frac{1}{Z^{1/2}}\sum_n e^{-\frac{\beta}{2}E_n}|n \rangle_L |n \rangle_R\,.
\end{equation}
This state displays a very atypical left-right pattern of entanglement at $t=0$ and the chaotic nature of the boundary theories is manifested by the fact that small perturbations added to the system in the asymptotic past destroy these delicate correlations \cite{BHchaos1}. This phenomenon is known as the butterfly effect.

An efficient way to diagnose this pattern of entanglement and how it is destroyed by small perturbations is to consider the mutual information $I(A,B)$ between subsystems $A\subset\text{QFT}_L$ and $B\subset\text{QFT}_R$, defined as
\begin{equation}\label{MIdef}
I(A,B)=S_A+S_B-S_{A \cup B}\,,
\end{equation}
where $S_A$ is the entanglement entropy of the subsystem $A$, and so on. Importantly, this quantity is always positive and provides an upper bound for correlations between operators $\mathcal{O}_L$ and $\mathcal{O}_R$ defined on $A$ and $B$, respectively \cite{boundIAB}
\be
I(A,B)\geq\frac{(\langle\mathcal{O}_L\mathcal{O}_R\rangle-\langle\mathcal{O}_L\rangle\langle\mathcal{O}_R\rangle)^2}{2\langle\mathcal{O}_L^2\rangle\langle\mathcal{O}_R^2\rangle}\,.
\ee

Let us consider a small perturbation by acting with an operator $W$ at some time $t_0$ in the past. From the point of view of the gravitational theory the state $W|TFD \rangle$ is represented by an excitation near the boundary of the space time, which then falls into the black hole. This excitation gets blue shifted as it fall into the black hole and generates a shock wave geometry, in which the wormhole becomes larger. If this perturbation is early enough, the operator will scramble the  Hilbert space and the state $W|TFD \rangle$ will have a zero mutual information between $A$ and $B$ at $t=0$, signaling the destruction of the left-right correlations. In this setup, then, the disruption of mutual information sets a bound on the two-sided correlators of the form
\be
\langle\mathcal{O}_L\mathcal{O}_R\rangle_W=\langle TFD|W^\dag\mathcal{O}_L\mathcal{O}_RW|TFD\rangle\,,
\ee
which are related by analytic continuation to the one-sided out-of-time-order correlators that appear in the chaos commutator (\ref{commutator}) \cite{BHchaos4}.
Therefore, the disruption of mutual information effectively provides a concrete realization of the butterfly effect in holographic theories \cite{BHchaos1}. This setup has been studied and extended in various directions in \cite{Leichenauer-2014,Sircar-2016,Ling-2016,Cai-2017,Qaemmaqami-2017,Wu-2017,Qaemmaqami-2017-2,Ahn-2017,Avila-2018}.

The disruption of the two-sided mutual information in the TFD state takes place at time scales of the order of the scrambling time $t_*\sim\beta\log N^2$, and is controlled by the so-called entanglement velocity $v_E$. Upon inspection one finds that the non-trivial part of the computation comes from the last term in (\ref{MIdef}) which, for large enough subsystems and times in the range $t_d <\!\!< t <\!\!< t_*$, is found to vary linearly with the shock wave time,
\be\label{vEdef}
\frac{d S_{A \cup B}}{dt_0} = v_E\, s_\mt{th}\, A_{\Sigma}\,,
\ee
where $s_\mt{th}$ is the thermal entropy and $A_{\Sigma}$ is the area of $\Sigma=\partial(A\cup B)$. This behavior can be explained in terms of the so-called `entanglement tsunami' that appears in the study of entanglement entropy following a quantum quench, both in field theory \cite{Calabrese:2005in} and holographic calculations \cite{AbajoArrastia:2010yt,Albash:2010mv,HM,tsunami1,tsunami2}. In \cite{tsunami1,tsunami2}, the authors conjectured the entanglement velocity should be bounded by
\be
v_E \leq v_E^\mt{Sch}=\frac{\sqrt{d}(d-1)^{\frac{1}{2}-\frac{1}{d}}}{\left[2(d-1) \right]^{1-\frac{1}{d}}}\,,
\ee
where $v_E^\mt{Sch}$ is the entanglement velocity for a $(d+1)$-dimensional Schwarzschild black brane. Later in \cite{Mezei-2016v2}, this bound was proven to be valid for quite generic holographic theories in Einstein gravity satisfying the NEC. However, once again, the bound was shown to be violated once the assumption of isotropy is relaxed \cite{Jahnke-2017}. In this case, though, $v_E$ is still bounded and never exceeds the speed of light.

More generally, \cite{Mezei-2016} conjectured that in any quantum system $v_E\leq v_B$. So, if the bound (\ref{boundvB2}) holds true, then, the entanglement velocity must also be bounded
\be\label{boundvE}
v_E\leq1\,.
\ee
The authors of \cite{Casini:2015zua} proved this using
the positivity of mutual information, while \cite{Hartman-2015} used inequalities of relative entropy.\footnote{Strictly speaking, the bound (\ref{boundvE}) holds true for large enough subsystems. For small subsystems, the `entanglement tsunami' picture breaks down and (\ref{boundvE}) can be violated instantaneously \cite{Kundu:2016cgh,Lokhande:2017jik}. However, causality still implies that in average, $v_E^{\text{avg}}<1$ throughout a unitary evolution.} However, both \cite{Casini:2015zua,Hartman-2015} assumed that the theory is Lorentz invariant. If the theory is Lorentz invariant the entanglement entropy depends not on the particular Cauchy slice, but on the causal development of the subregion. This means that one can split the Hilbert space in various ways (basically we can pick any space-like slice) and the entanglement entropy of boosted regions is trivially related. Various consequences follows from it, such as the entropic proof of the c-theorem \cite{Casini:2004bw,Casini:2006es}, monotonicity properties of various entanglement related quantities, e.g. \cite{Myers:2010tj,Myers:2012ed,Casini:2012ei}, and so on. Since we are studying a system that does not have Lorentz invariance, these proofs do not apply and, in particular, we do not expect the speed of light to play a role. See for instance \cite{Kusuki:2017jxh} for a discussion of entanglement entropy on generic time slices for theories that are not Lorentz invariant.

\subsection{Plan of the paper}

The paper is organized as follows. In section \ref{sec-prelims} we give a brief overview of the background material needed to set up the problem. We introduce and explain the basic properties of the $\mathcal{N}=4$ non-commutative super Yang Mills theory and its gravity dual, and then we discuss some subtleties in the definition of the gauge invariant observables of interest, namely, correlation functions and entanglement entropies. In section \ref{sec-gravity} we explain how to construct shock wave solutions with definite momentum for a very general two-sided black hole geometry, including geometries which are not asymptotically AdS. Then, we show how to extract from the shock wave profiles several chaotic quantities of interest: the maximal Lyapunov exponent, the scrambling time and the butterfly velocity. We specialize our formulas to the gravity dual of $\mathcal{N}=4$ non-commutative super Yang Mills. In section \ref{sec-IAB} we compute the two-sided mutual information for strip-like regions both in the unperturbed geometry and in the prepense of homogeneous shock waves. We also discuss the role of the spread of entanglement in the disruption of the two-sided mutual information in the shock wave geometries. In section \ref{sec-VB} we discuss an alternative derivation of the butterfly velocity, based on entanglement wedge subregion duality, and we show that the final result agrees with the shock wave calculations. Finally,  we close in section \ref{sec-discussion} with a discussion of our results and open questions.  We relegate some technical details to the appendices.

\section{Preliminaries}\label{sec-prelims}

\subsection{Gravity dual of non-commutative SYM}

Non-commutative quantum field theory has been an important theoretical arena and a topic of great research interest in the past few decades.
The basic postulate of non-commutativity is that space-time coordinates do not commute. Instead, they satisfy the following commutation relation
\be\label{commreltheta}
\left[x^\mu,x^\nu\right]=i\theta^{\mu\nu}\,,
\ee
where $\theta^{\mu\nu}$ is a real and antisymmetric rank-2 tensor. The algebra of functions in a non-commutative theory can be viewed as an algebra of
ordinary functions with the product deformed to the so-called Moyal product,
\be
\left(\phi_1\star\phi_2\right)(x)\equiv e^{\frac{i}{2}\theta^{\mu\nu}\partial_\mu^y\partial_\nu^z}\phi_1(y)\phi_2(z)\big|_{y=z=x}\,.
\ee

Non-commutative theories arise naturally in string theory, as the worldvolume theory of D-branes with non-zero NS-NS $B$-field, provided that one takes a special limit to decouple the open and closed string sectors \cite{Douglas:1997fm, Ardalan:1998ce,Seiberg:1999vs,Alishahiha:1999ci}. In the context of gauge/gravity duality, this implies that the dynamics of certain strongly-coupled, large $N$, non-commutative field theories can be described in terms of a classical gravity dual. The first example of this kind of dualities was presented in \cite{Hashimoto:1999ut,Maldacena-Russo}, which studied a specific decoupling limit of a stack of D3-branes with non-zero $B_{23}$. The decoupling limit consists of scaling the string tension to infinity, and the closed string metric to zero, while keeping the $B$-field fixed. This limit provided a gravity dual for finite temperature $SU(N)$ non-commutative super Yang Mills theory at large $N$ and large 't Hooft coupling $\lambda$, with non-commutative parameter non-zero only in the $(x^2,x^3)$-plane, i.e., $[x^2,x^3]\sim i \theta$. The gravity dual of this theory is type IIB supergravity, with
\begin{eqnarray}\label{backg}
ds_\mt{E}^2 &=&\frac{ R^2}{{\hat g}^{1/2}h(r)^{1/4}} \left[r^2\left(-f(r) dt^2+dx_1^2+h(r)\left(dx_2^2+dx_3^2\right)\right)+ \frac{dr^2}{r^2f(r)}+ d\mathrm{\Omega}_5^2\right],\nonumber\\
e^{2\Phi} &=& {\hat g}^2h(r)\,,\nonumber\\
B_{23} &=& R^2a^2 r^4 h(r)\,,\\
C_{01} &=& \frac{R^2a^2}{\hat g}r^4\,,\nonumber\\
F_{0123r} &=& \frac{4R^4}{{\hat g}}r^3 h(r)\,,\nonumber
\end{eqnarray}
where $R^4 = 4 \pi \hat g N\alpha'^2$, $\hat g$ denotes the string coupling and $\alpha'$ is the string tension. The 't Hooft coupling is related to the curvature of the background and the string tension through the standard relation $\sqrt{\lambda}=R^2/\alpha'$. Notice that for future convenience, we have given the metric above in the Einstein frame.\footnote{In the string frame $ds_\mt{str}^2=e^{\Phi/2}ds_\mt{E}^2$.}
Moreover,
\be
f(r)=1-\frac{\rh^4}{r^4}
\ee
is the standard blackening factor, with $\rh=\pi T$, while
\be
h(r)=\frac{1}{1+a^4r^4}
\ee
is a function that encodes the effects of the non-commutativity.\footnote{A simple way to understand why this background is dual to a non-commutative theory is to
consider an open string in the corresponding background, which yields the commutation relation (\ref{commreltheta}) \cite{Seiberg:1999vs}.} The parameter $a$ is related to the non-commutative parameter $\theta$ through $a=\lambda^{1/4}\sqrt{\theta}$. This parameter can be thought of as a ``renormalized'' non-commutative length scale at strong coupling, since this is the parameter that will enter in every holographic computation.

As usual in holography, the radial direction $r$ is mapped into an energy scale in the field theory, in such a way that $r\rightarrow\infty$ and $r\rightarrow \rh$ correspond to the UV and IR limits, respectively. The directions $x^\mu\equiv(t,\vec{x})$ are parallel to the boundary and are directly identified with the field theory directions. Finally, the five-sphere coordinates are associated with the global $SU(4)$ internal symmetry group, but they will play no role in our discussion.

For $r-\rh <\!\!< a^{-1}$, the background \eqref{backg} goes over to the AdS$_5\text{-Schwarzschild} \times {\rm S}^5$ solution, which is dual to a thermal state of the standard $SU(N)$ super Yang Mills theory. Indeed, it can be shown that all the thermodynamic quantities derived from \eqref{backg} are the same as the ones obtained from the AdS$_5$-Schwarzschild solution \cite{Hashimoto:1999ut,Maldacena-Russo}. This observation just reflects the fact that the non-commutative boundary theory goes over to ordinary super Yang Mills at length scales much greater than $\lambda^{1/4}\sqrt{\theta}$. On the other hand, for $r>\!\!> a^{-1}$ the background \eqref{backg} exhibits significant differences with respect to AdS$_5 \times {\rm S}^5$ and, in particular, is no longer asymptotically AdS. From the boundary perspective, this just means that the effect of the non-commutativity becomes pronounced for length scales of order or smaller than $\lambda^{1/4}\sqrt{\theta}$.

\subsection{Gauge invariant operators in non-commutative theories}\label{sec-gaugeinv}

In non-commutative gauge theories, the non-commutativity of the spacetime mixes with the gauge transformations and, therefore, there are no gauge invariant operators in position space. However, one can construct gauge invariant operators in momentum space, $\tilde{{\cal O}} (k^\mu)$, by smearing the gauge covariant operators ${\cal O}(x^\mu)$ transforming in the adjoint representation of the gauge group over an open Wilson line $W(x,C)$ according to \cite{Ishibashi:1999hs,Das:2000md,Gross:2000ba}
\begin{align}\label{wline}
\tilde{{\cal O}}(k) = \int d^4x\,{\cal O} (x)\star W(x,C) \star e^{i k \cdot x}\,,
\end{align}
where $\star$ denotes the Moyal product. A few comments are in order:
\begin{itemize}
  \item For $k\sqrt{\theta}<\!\!<1$, the length of the Wilson line $\ell_W$ goes to zero and (\ref{wline}) reduces to the standard operators in commutative field theory,
  \be
  \tilde{{\cal O}}(k)\to{\cal O}(k)=\int d^4x\,  {\cal O} (x)e^{i k \cdot x}\,.
  \ee
  \item For $k\sqrt{\theta}>\!\!>1$, the length of the Wilson line $\ell_W$ becomes large as dictated by the non-commutativity. In this limit, the operator is dominated by the Wilson line regardless of what operator is attached at the end. Therefore, the correlation functions of these operators are expected to exhibit a universal behavior at large $k\sqrt{\theta}$. One concrete example of this fact is the universal dissipation time at large momentum found in the quasinormal mode analysis of \cite{Edalati:2012jj}.
  \item Finally, the fact that $\tilde{{\cal O}}(k)$ contains a Wilson line whose length $\ell_W \simeq \theta k $ depends on $k$
implies that one should not think of these operators as the ``same'' operator ${\cal O}(x)$ with different momentum as we usually
do in standard field theory. In particular, one should not expect to obtain a local operator by Fourier
transforming to position space. Instead, one should think of $\tilde{{\cal O}}(k)$ as genuinely different operators at
different $k$.
\end{itemize}

In the holographic context, there is a one-to-one map between gauge invariant operators ${\cal O}(x)$ in the boundary theory and local bulk fields $\varphi(r,x)$.  According to the standard dictionary, the non-normalizable mode of $\varphi(r\to\infty,x)$ in a near-boundary expansion corresponds to the source of the dual operator ${\cal O}(x)$ while the normalizable mode gives its expectation value. The above map is subtle when the boundary gauge theory is non-commutative because, as explained above, there are no gauge invariant local operators in position space. This issue is solved by working in momentum space. More specifically, one can assume that the bulk field $\varphi(r,k)$ is dual to a gauge invariant operator $\tilde{{\cal O}}(k)$ of the form \eqref{wline} in the sense that in the boundary theory there is a coupling of the form
\begin{align}\label{BoundaryCoupling}
S=S_0+\int d^4k\,\varphi_0(-k)\tilde{{\cal O}}(k)\,.
\end{align}
As usual, the source $\varphi_0(k)$ is determined from the non-normalizable mode of $\varphi(r,k)$ given some appropriate boundary condition in the IR.

\subsection{Out-of-time-ordered correlators in momentum space}

In order to diagnose chaos we need to compute the norm of the commutator $C(t,\vec{x})$, defined in (\ref{eq-C(t,x)}), for two Hermitian gauge-invariant operators
$W(t,\vec{x})$ and $V(0,0)$. As expected, such definition is problematic for non-commutative gauge theories, because in these theories there are no gauge invariant operators in position space. Instead, we will quantify chaos in non-commutative theories by defining an equivalent quantity in momentum space, i.e.,
\begin{equation}
C(t,\vec{k})=- \langle [W(t,\vec{k}),V(0,0)]^2 \rangle\,.
\end{equation}
There is no need to go to frequency space, since the non-commutativity only acts on the spatial coordinates. In the next section we show that $C(t,\vec{k})$ has a pole precisely at $|\vec{k}|=i \lambda_L /v_B$, from which we can extract the Lyapunov exponent $\lambda_L$ and the butterfly velocity $v_B$. The fact that the pole of $C(t,\vec{k})$ gives the Lyapunov exponent and the butterfly velocity is implicit in other holographic calculations. See for instance the Appendix C of \cite{Gu:2016oyy}.

\subsection{Entanglement entropy in non-commutative theories} \label{sec-EE-nc}

In standard quantum field theory the entanglement entropy associated to a subsystem $A$ can be calculated by the von Neumann formula, $S_A=-\text{tr}\left( \rho_A \log \rho_A \right)$, where $\rho_A$ is the reduced density matrix associated to $A$. In holographic theories $S_A$ can be computed in the bulk by the HRRT prescription \cite{RT,HRT}
\be
S_A=\frac{\text{Area}(\gamma_A)}{4 G_\mt{N}}\,,
\label{eq-RT}
\ee
where $\gamma_A$ is an extremal area surface whose boundary coincides with the boundary of the region $A$, i.e., $\partial\gamma_A=\partial A$. Entanglement entropy in local theories follows the so-called {\it area law}, which means that the leading UV divergence of $S_A$ has a coefficient which is proportional to the area of the boundary of the region $\Sigma=\partial A$,
\be\label{arealaw}
S_A\sim\frac{A_\Sigma}{\epsilon^{d-2}}+\cdots\,.
\ee
The area law basically means that the entanglement between $A$ and its complement $\bar{A}$ is dominated by contributions coming from short-ranged interactions between points close to the boundary between $A$ and $\bar{A}$.

In non-commutative theories it is not always possible to precisely define the curve (or surface) delimiting the region $A$. One possible way to define a subsystem in these theories was proposed in \cite{Fischler-2013}. The prescription is the following: first, one defines a region $A$ for the commutative case as
\be
A=\{(x_1,x_2,x_3)\,\, \text{such}\,\,\text{that}\,\,\Phi(x_1,x_2,x_3) \leq 0 \}\,,
\ee
where the surface $\Phi(x_1,x_2,x_3) = 0$ defines the boundary of the region $A$. Then, one promotes $\Phi$ to an operator,
\be
\Phi(x_1,x_2,x_3) \rightarrow \hat{\Phi}(\hat{x}_1,\hat{x}_2,\hat{x}_3)\,.
\ee
Let $| \Phi \rangle$ denote the eingenvector of $\hat{\Phi}$, with eigenvalue $\Phi$, i.e.,
\be
\hat{\Phi} |\Phi \rangle = \Phi |\Phi \rangle\,.
\ee
The subsystem $A$ can then be uniquely defined as
\be
A = \{  |\Phi \rangle \,\, \text{such}\,\,\text{that}\,\,\Phi \leq 0 \}\,.
\label{eq-regionA}
\ee

For holographic theories $S_A$ can still be computed by using the standard HRRT prescription (\ref{eq-RT}), where the boundary of $A$ is given be the classical entangling surface at a particular cut-off scale \cite{Fischler-2013,Karczmarek:2013xxa}. With this holographic definition, it has been shown that for small enough regions the entanglement entropy follows instead a {\it volume law},
\be
S_A\sim\frac{V_A}{\epsilon^{d-1}}+\cdots\,,
\ee
while for large regions the standard area law (\ref{arealaw}) is recovered  \cite{Fischler-2013,Karczmarek:2013xxa}. This transition from a volume law to an area law behaviour has also been observed in quantum field theory calculations \cite{Karczmarek:2013jca,Sabella-Garnier:2014fda,Okuno:2015kuc} and has been understood as a result of the non-locality inherent of non-commutative theories \cite{Shiba:2013jja,Kol:2014nqa}. Very recently, the full cutoff dependence was studied in \cite{Chen:2017kfj} which found an exact match with respect to the results previously obtained in the strong coupling regime through holography.

Finally, we point out that the time dependence of entanglement entropy for a free scalar field on a non-commutative sphere following a quantum quench was studied in \cite{Sabella-Garnier:2017svs}. In this paper it was found that the entanglement velocity $v_E$ is generically larger that the commutative counterpart, even exceeding the speed of light in the limit of very strong non-commutativity. As explained in the introduction, this is not an issue for non-commutative theories since Lorentz invariance is explicitly broken and the standard notions of causality do not apply. However, this raises a number of questions. Does this behavior hold in the strong coupling regime? And more importantly, does the conjecture that $v_E\leq v_B$ \cite{Mezei-2016} holds for general non-local theories? If so, what are the implications for the transfer of quantum information?

\section{Perturbations of the TFD state} \label{sec-gravity}

\subsection{Eternal black brane geometry}
Let us consider a two-sided black brane geometry of the form
\begin{equation}
ds^2 = G_{MN}(r)dx^M dx^N=-G_{tt}(r)dt^2+G_{rr}(r)dr^2+G_{ij}(r)dx^i dx^j\,,
\label{eq-metric0}
\end{equation}
where $i,j=1,2,...9$. Here $(t,x^i)$ with $i=1,2,3$ are the coordinates of the boundary theory, while  $x^i$ with $i=4,5,...,9$ are the coordinates on the $S^5$. Also note that $r$ denotes the holographic radial coordinate. We take the boundary to be located at $r= \infty$ and the horizon at $r=\rh$. We assume the following near-horizon expressions for the metric functions
\begin{equation}
G_{tt}=c_0 (r-\rh)\,,\,\,\,\,\,\,\,\,G_{rr}=\frac{c_1}{r-\rh}\,,\,\,\,\,\,\,\,\,G_{ij}(\rh)=\text{constant}.
\label{eq-nearH}
\end{equation}
The inverse Hawking temperature associated to the above metric is
\begin{equation}
\beta \equiv \frac{1}{T}=4 \pi\sqrt{\frac{c_1}{c_0}}\,.
\end{equation}
In the study of shock waves is convenient to work in Kruskal coordinates, since these coordinates cover smoothly the two sides of the geometry. We first define the Tortoise coordinate
\begin{equation}
d r_*=\sqrt{\frac{G_{rr}}{G_{tt}}} dr\,,
\end{equation}
and then we introduce the Kruskal coordinates $U,V$ as follows,
\begin{equation}
UV=e^{\frac{4\pi}{\beta}r_*}\,,\,\,\,\,\,\,\,\,U/V = - e^{-\frac{4\pi}{\beta}t}\,.
\end{equation}
In terms of these coordinates the metric reads
\begin{equation}
ds^2=2A(U,V)dU dV+G_{ij}(U,V)dx^i dx^j\,,
\end{equation}
where
\begin{equation}
A(U,V)=\frac{\beta^2}{8\pi^2}\frac{G_{tt}(U,V)}{UV}\,.
\label{eq-metric0Kruskal}
\end{equation}
The region $U>0$ and $V<0$ ($U<0$ and $V>0$) covers the left (right) exterior region, while the region $U>0$ and $V>0$ ($U<0$ and $V<0$) covers the black hole (white hole) interior region. The horizon is located at $UV=0$. The boundary (left or right) is located at $UV=-1$ and the singularity at $UV=1$. We assume that the unperturbed metric is a solution of Einstein's equations\footnote{A possible cosmological constant term is absorbed into the definition of the stress-energy tensor.}
\begin{equation}
R_{MN}-\frac{1}{2}G_{MN}R= 8 \pi G_N T_{0MN}^\text{matter}\,,
\label{eq-EoMunperturbed}
\end{equation}
where the stress-energy tensor is assumed to be of the form
\begin{equation}
T_{0}^\text{matter}= T_{MN}dx^M dx^N=2\,T_{UV} dUdV+T_{UU}dU^2+T_{VV}dV^2+T_{ij}dx^i dx^j\,.
\end{equation}
$T_{MN}=T_{MN}(U,V,x^i)$ is the most general stress-energy tensor which is consistent with the Ricci tensor of the unperturbed geometry.

\subsection{Shock wave geometries}

In this section we study how the metric (\ref{eq-metric0}) changes when we add to the system a null pulse of energy located at $U=0$ and moving in the $V$-direction. The motivation to consider such a perturbation is the following. In the context of gauge/gravity duality the two-sided black brane geometry is dual to a thermofield double state of two copies of the boundary theory. This thermofield double state has a very particular pattern of entanglement between the two boundaries theories at $t=0$. We want to know how this pattern of entanglement changes when we perturb one of the boundary theories far in the past. We can do that by inserting an operator in one of the boundary theories at some time $t_0$ in the past and studying the evolution of the system. In the gravitational description, this corresponds to the creation of a perturbation close to the boundary, which then falls into the black brane. From the point of view of the $t=0$ frame, the energy of this perturbation increases exponentially with $t_0$, while its distance from the past horizon decreases exponentially with $t_0$. As a results, an early enough perturbation will follow an almost null trajectory very close to the past horizon, which can then replaced by a null pulse of energy, located at $U=0$ and moving in the $V$-direction. This pulse of energy will give rise to a shock wave geometry.

\subsubsection{Shocks with definite momentum} \label{sec-VB-NC}

In non-commutative theories is not possible to define local gauge invariant operators in position space. However, as explained in section \ref{sec-gaugeinv}, one can define gauge invariant operators that are local in momentum space. This is done by smearing a gauge covariant operator $\mathcal{O}(x)$ over a Wilson line. Despite being non-local, this type of perturbation can also give rise to a shock wave geometry, specifically, a shock wave geometry with definite momentum $k$. The only requirement is that the perturbation is local in time and is applied in the asymptotic past. This is perfectly possible in non-commutative SYM theory, since the non-commutativity affects only two spatial coordinates, i.e. $[x_2,x_3]=i \theta$.

Based on this observation, we will consider the following form for the stress-energy tensor of the shock wave,
\begin{equation}
T_{UU}^\text{shock}=E\, e^{\frac{2 \pi}{\beta}t} \delta(U) e^{i \vec k \cdot \vec x}\,.
\label{eq-Tshock}
\end{equation}
where $k^i =0$ for $i=4,5...,9$.
This corresponds to a pulse of energy of definite momentum $\vec k$ (along the boundary coordinates) and amplitude $E$. The pulse world line divides the bulk into two regions, the causal future of the pulse (region $U>0$), and its causal past (region $U<0$), but only the metric in the causal future of the pulse gets modified by its presence. The metric in the causal past, on the other hand, is the same as the unperturbed metric.

It turns out that the backreaction of this pulse of energy is very simple. It can be described by a shift $V \rightarrow V + \alpha$ in the $V$-coordinate \cite{Dray-85,Sfetsos-94}, where $\alpha$ can be determined from Einstein's equations, as we explain below. Given the form of the stress-energy tensor (\ref{eq-Tshock}), $\alpha$ should take the following form,
\be
\alpha=\tilde{\alpha}(t,\vec k) e^{i \vec k \cdot \vec x}\,.
\label{eq-alphaAnsatz}
\ee
We will now use Einstein's equations to determine $\tilde{\alpha}(t,\vec k)$.

We start by replacing $V$ by $V+\Theta(U) \alpha$ in the unperturbed metric (\ref{eq-metric0}). Note that the Heaviside step function $\Theta(U)$ guarantees that only the causal future of the pulse ($U>0$) is affected by its presence. The shock wave geometry can then be written as
\begin{equation}
ds^2= 2A(U,V+\Theta \alpha)dU (dV+i\Theta k_i \alpha\, dx^i)+G_{ij}(U,V+\Theta \alpha)dx^i dx^j\,,
\end{equation}
while the stress-energy tensor reads
\be
\begin{split}
T^\text{matter}=&\,2\,T_{UV}(U,V+\Theta \alpha) dU (dV+i\Theta\, k_i \alpha\, dx^i)+T_{UU}(U,V+\Theta \alpha)dU^2\\
&+T_{VV}(U,V+\Theta \alpha)(dV+i\Theta\, k_i \alpha\, dx^i)^2+T_{ij}(U,V+\Theta \alpha)dx^i dx^j\,,
\end{split}
\ee
For simplicity, we define the new coordinates
\begin{equation}
\hat{U}=U\,,\,\,\,\,\,\,\hat{V}=V+\Theta \, \alpha\,,\,\,\,\,\,\, \hat{x}^i=x^i\,,
\label{eq-hatcoordinates}
\end{equation}
in which terms the metric and the stress-energy tensor take the form
\begin{equation}
ds^2=2\hat{A}\, d\hat{U} d\hat{V}+\hat{G}_{ij}\,d\hat{x}^i d\hat{x}^j-2 \hat{A} \,\hat{\alpha}\, \delta(\hat{U}) \,d\hat{U}^2\,,
\label{eq-metricAnsatz}
\end{equation}
and
\be
\begin{split}
T^\text{matter}=&\,2\left[ \hat{T}_{\hat{U}\hat{V}}-T_{\hat{V}\hat{V}}\,\hat{\alpha}\,\delta(\hat{U}) \right] d\hat{U} d\hat{V}+\hat{T}_{\hat{V}\hat{V}} d\hat{V}^2+\hat{T}_{ij} d\hat{x}^i d\hat{x}^j\\
&+\left[\hat{T}_{\hat{U}\hat{U}}+\hat{T}_{\hat{V}\hat{V}}\,\hat{\alpha}^2 \delta(\hat{U})^2-2 \hat{T}_{\hat{U} \hat{V}}\,\hat{\alpha}\,\delta(\hat{U}) \right]d\hat{U}^2\,,
\label{eq-Tmatter}
\end{split}
\ee
respectively. The hats in these expressions indicate that the corresponding quantities are evaluated at $(\hat{U},\hat{V},\hat{x^i})$.
Finally, we determine $\alpha$ by requiring (\ref{eq-metricAnsatz}) to satisfy the Einstein's equations
\begin{equation}
R_{MN}-\frac{1}{2}G_{MN}R= 8 \pi G_N \left( T_{MN}^\text{matter} +T_{MN}^\text{shock}\right)\,,
\label{eq-EoMshock}
\end{equation}
with $T^\text{shock}$ and $T^\text{matter}$ given by (\ref{eq-Tshock}) and (\ref{eq-Tmatter}), respectively. In order to simplify the notation, in the following we will drop the hat over the symbols, but keeping in mind that we are really dealing with the coordinates defined in (\ref{eq-hatcoordinates}).

The analysis of the equations of motion simplifies when we rescale $\alpha$ and $T^\text{shock}$ as $\alpha \rightarrow \epsilon \alpha$ and $T^\text{shock} \rightarrow \epsilon T^\text{shock}$. With this rescaling we can recover the equations of motion (\ref{eq-EoMunperturbed}) for the unperturbed metric by setting $\epsilon =0$ in (\ref{eq-EoMshock}). Furthermore, by using (\ref{eq-EoMunperturbed}) and analyzing the terms proportional to $\epsilon$ in (\ref{eq-EoMshock}) we find that $\tilde{\alpha}$ needs to satisfy the equation\footnote{To obtain this equation we use that $\delta'(U)G_{ij,V}=-\delta(U)G_{ij,UV}$ and $U^2 \delta(U)^2 =0$.}
\begin{equation}
\delta(U) G^{ij} \left(-A\,k_i k_j+\frac{1}{2}G_{ij,UV} \right)\tilde{\alpha} = 8 \pi T_{UU}^\text{shock}\,.
\label{eq-alphaTuu}
\end{equation}
Going back to the original coordinates $t$ and $r$ of metric (\ref{eq-metric0}), the equation for $\tilde{\alpha}$ reads
\begin{equation}
\left( G^{ij}k_i k_j+M^2 \right)\tilde{\alpha}(t,\vec k)= - e^{2 \pi (t-t_*) / \beta} \,,
\label{eq-alpha-x}
\end{equation}
where
\be\label{scrambt}
t_*=\frac{\beta}{2 \pi} \log \frac{A(\rh)}{8 \pi G_\mt{N} E}\,.
\ee
and
\be
M^2=\left( \frac{2\pi}{\beta} \right)^2 \frac{G^{ii}(\rh)G_{ii}'(\rh)}{G_{tt}'(\rh)}=  \left( \frac{2\pi}{\beta} \right)^2 \frac{1}{G_{tt}'(\rh)} \left[ \frac{G_{11}'}{G_{11}}+2\frac{G_{22}'}{G_{22}}+5 \frac{G_{\theta \theta}'}{G_{\theta \theta}}\right]\,.
\ee
Notice that we have written the metric on the $S^5$ as $G_{\theta \theta}(r)\, d \Omega_5^2$.
Assuming $G_{ij}$ to be diagonal, the shock wave profile is then given by
\be\label{eq-alpha-k}
\tilde{\alpha}(t,\vec{k}) = \frac{ e^{\frac{2\pi}{\beta} (t-t_*)}}{G^{ii}(\rh)k_i^2+M^2}\,.
\ee

\subsubsection{Lyapunov exponent and scrambling time}
We can extract the chaotic properties of the boundary theory by identifying $\tilde{\alpha}(t,\vec k)$ with $C(t,\vec k) = -\langle [W(t,\vec k),V(0,0)]^2\rangle$.
By setting $\vec{k}=0$ we obtain
 \be
 \tilde{\alpha} = \text{constant}\times e^{\frac{2 \pi}{\beta}(t-t_*)}\,,
 \label{eq-alphaConstant}
 \ee
where the constant of proportionality is of order $\mathcal{O}(1)$. This case corresponds to a homogeneous shock wave geometry. From this profile we can readily extract the Lyapunov exponent
\be
\lambda_L=\frac{2\pi}{\beta}\,,
\ee
and the scrambling time $t_*$, given in (\ref{scrambt}). Using the expression for the Bekenstein-Hawking entropy $S_\mt{BH}=A(\rh)/4G_\mt{N}$, we can write the leading order contribution to the scrambling time as
\be
t_*= \frac{\beta}{2\pi} \log S_\mt{BH}\,.
\ee
Note that, since $\beta$ and $S_\mt{BH}$ are not affected by the non-commutative parameter, both $\lambda_L$ and $t_*$ are precisely the same as for the commutative version of the SYM theory.

Later in section \ref{sec-disruptMI}, we will use this type of shock waves to study the disruption of two-sided mutual information. From this study we will extract another quantity of interest, the so-called entanglement velocity $v_E$.

\subsubsection{Butterfly velocity}
At finite $\vec{k}$, we expect the size of the Wilson line coupled to the operator $W(t,\vec k)$ to be small for $\sqrt{\theta}k<\!\!<1$ and large for $\sqrt{\theta}k>\!\!>1$. Hence, in the limit of low momentum one can expect to recover an approximate exponential behaviour as in (\ref{eq-C(t,x)}), for $x>\!\!>\sqrt{\theta}$ and $t>\!\!>t_d$. More generally, we can extract $v_B$ from the leading pole of $C(t,\vec k)$. More specifically, it can be shown that $\tilde{\alpha}(t,\vec{k})$ has a pole precisely at\footnote{This seems to be consistent with the general hydrodynamic theory of quantum chaos proposed in \cite{Blake-2017}. In that paper the authors propose that the behaviour of OTOCs are controlled by a hydrodynamic chaos mode $\sigma_\mt{hydro}(k)$, which also has a pole precisely at $|\vec k| = i \lambda_L/v_B$.}
\be
|\vec k|=\sqrt{k_1^2+k_2^2+k_3^3}=i \frac{\lambda_L}{v_B(\phi)}\,.
\ee
We refer the reader to appendix \ref{appA0} for details. From (\ref{eq-alpha-k}), then, this leads to
\be
v_B^2(\phi)= \frac{G_{tt}'}{\left( 2\frac{G_{22}'}{G_{22}}+\frac{G_{11}'}{G_{11}} +5 \frac{G_{\theta \theta}'}{G_{\theta \theta}}\right)}\frac{1}{G_{22} \sin^2 \phi+G_{11} \cos^2 \phi} \Big|_{r=\rh}\,.
\label{eq-VBani}
\ee
In this formula $\phi$ is the angle between the $x_1-$direction and $\vec x$. The dependence of $v_B$ on $\phi$ is due to the anisotropy of the system ($G_{11}\neq G_{22}$), however, it is clear that in the limit of vanishing non-commutativity  ($G_{11}= G_{22}$) the $\phi$-dependence disappears.

\begin{figure}[t!]
\begin{center}
\setlength{\unitlength}{1cm}
\includegraphics[width=0.6\linewidth]{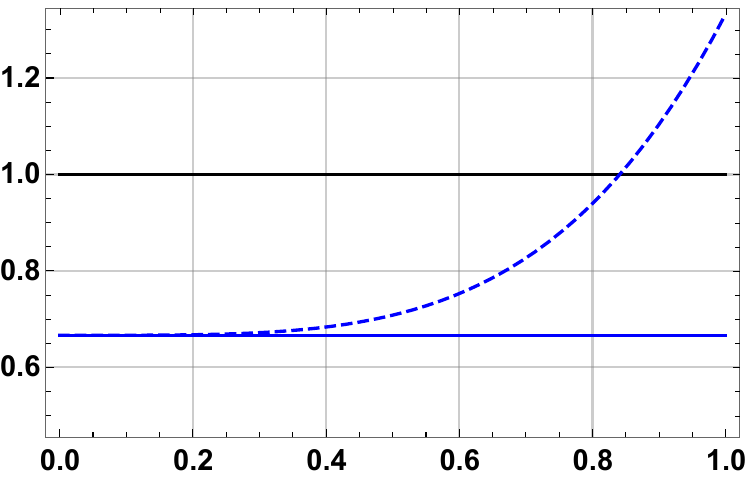}
\put(-4.5,-.3){\large $a \rh$}
\put(-9.75,+3.5){\large $v_{B}^2$}
\put(-2,+2){\large $v_{B,x_1}^2$}
\put(-2,+4.6){\large $v_{B,x_2}^2$}

\end{center}
\caption{ \small
Butterfly velocity squared $v_\mt{B}^2$ versus the dimensionless parameter $a \rh$. The horizontal blue line represents the butterfly velocity along the $x_1$-direction, which does not depend on $a \rh$ and takes the conformal value $v_\mt{B}^2=2/3$, while the dashed blue curve represents the butterfly velocity along the $x_2$- and $x_3$-directions. The horizontal black line corresponds to the speed of light.}
\label{fig-VB2}
\end{figure}

For later convenience, we write the explicit formulas for the butterfly velocity along the $x_1$-direction
\be
v_{B,x_1}^2 \equiv v_B^2(\phi=0) =\frac{G_{tt}'}{G_{11}\left(2\frac{G_{22}'}{G_{22}}+\frac{G_{11}'}{G_{11}} +5 \frac{G_{\theta \theta}'}{G_{\theta \theta}}\right)}\Big{|}_{r=\rh}\,,
\label{VBpara}
 \ee
and for the butterfly velocity along $x_2$- and $x_3$-directions
\be
v_{B,x_2}^2\equiv v_B^2(\phi=\pi/2)=\frac{G_{tt}'}{G_{22}\left(2\frac{G_{22}'}{G_{22}}+\frac{G_{11}'}{G_{11}}+5 \frac{G_{\theta \theta}'}{G_{\theta \theta}} \right)}\Big{|}_{r=\rh}\,.
\label{VBperp}
\ee
Specializing these formulas to the gravity dual of non-commutative SYM, whose metric is given by (\ref{backg}), we get
\be
v_B^2(\phi=0)=v_{B,x_1}^2=\frac{2}{3}\,,
 \ee
and
\be
v_B^2(\phi=\pi/2)=v_{B,x_2}^2=\frac{2}{3} (1+a^4\rh^4)\,.
 \ee
In figure \ref{fig-VB2} we plot $v_{B,x_1}^2$ and $v_{B,x_2}^2$ as a function of the non-commutative parameter $a \rh$. The component of the butterfly velocity along the commutative $x_1$-direction does not depend on $a \rh$ and takes the conformal value, i.e., $v_{B,x_1}^2=2/3$, while the the components along the non-commutative plane approach the conformal value only in the IR, $v_{B,x_2}^2(a \rh\to0)\to2/3$, and grow monotonically as $a \rh$ is increased.
It is interesting to note that $v_{B,x_2}^2$ exceeds the speed of light in the regime of strong non-commutativity. As explained in the introduction,  this is not an issue for non-commutative theories since Lorentz invariance is explicitly broken and the standard notions of causality do not apply. Nevertheless, this result is remarkable in the context of quantum information theory, since it represents a novel violation of the known bounds on the rate of transfer of information. We will comment more on this result in the conclusions.

Finally, it is instructive to discuss the physical meaning of $v_B$ in non-commutative theories. In the commutative case, the butterfly velocity describes the spatial growth of an operator around a point $\vec x$ where the operator is inserted. In the case of non-commutative theories, the operator is smeared over a Wilson line. Here $v_B$ presumably describes the growth of this operator around the curve $C$. We will confirm this intuition in section \ref{sec:vB2}.

\section{Entanglement velocity from two-sided perturbations} \label{sec-IAB}

In this section we compute the two-sided mutual information both in the unperturbed geometry and in the presence of a shock wave. As explained in the introduction, the disruption of the mutual information in the second case characterizes the butterfly effect in holographic theories.

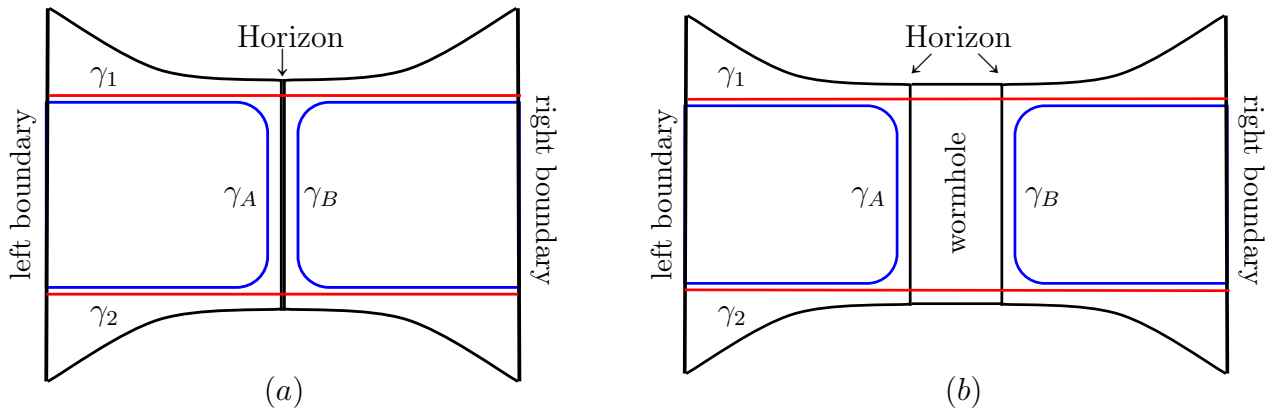
\begin{figure}[t!]
\begin{center}
\begin{tabular}{cc}
\setlength{\unitlength}{1cm}
\hspace{-0.9cm}
\includegraphics[width=7cm]{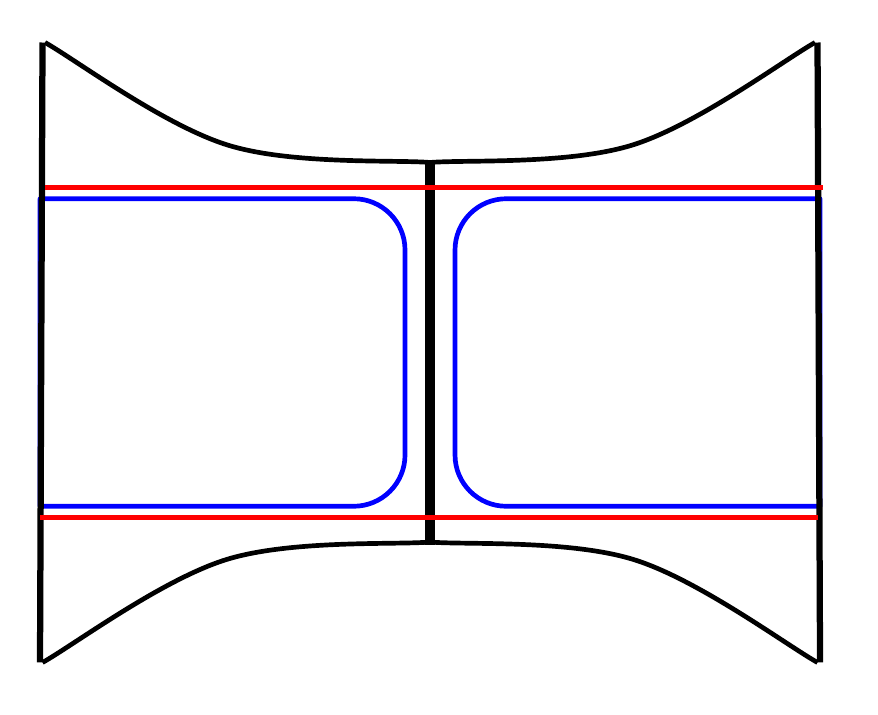}
\qquad\qquad &
\includegraphics[width=7.8cm]{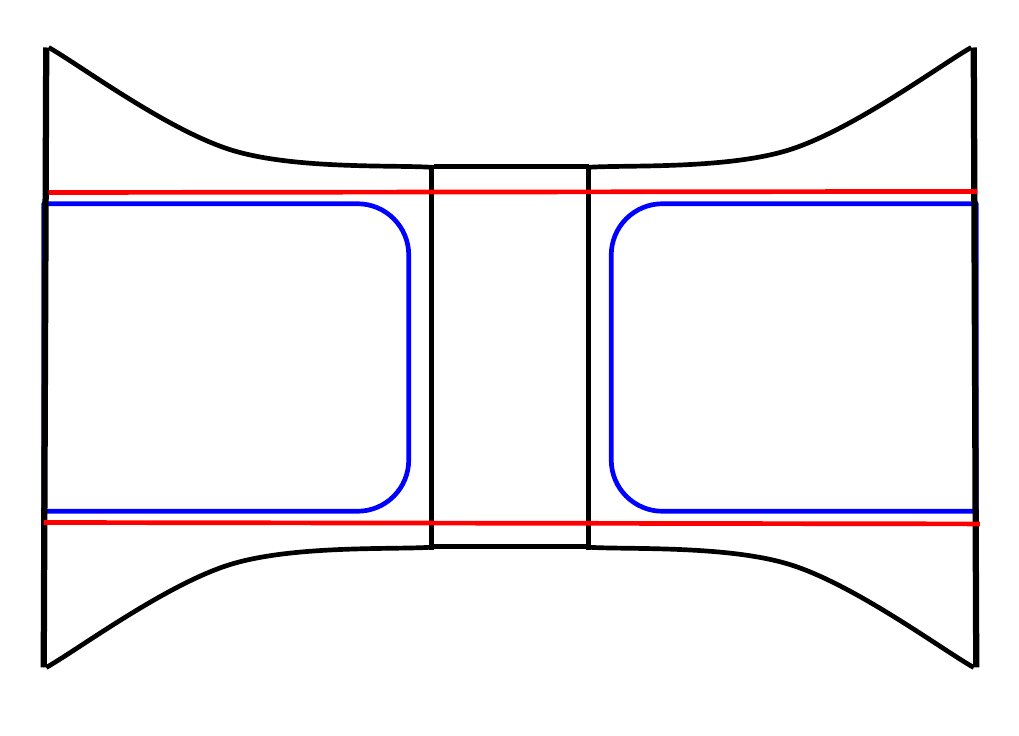}
\qquad

         \put(-385,80){$\gamma_A$}
         \put(-355,80){$\gamma_B$}

         \put(-150,80){$\gamma_A$}
         \put(-85,80){$\gamma_B$}

         \put(-435,125){$\gamma_1$}
         \put(-435,35){$\gamma_2$}

         \put(-200,125){$\gamma_1$}
         \put(-200,35){$\gamma_2$}

         \put(-380,138){$\text{Horizon}$}
         \put(-130,138){$\text{Horizon}$}

         \put(-115,5){$(b)$}
         \put(-370,5){$(a)$}

         \put(-366,138){\rotatebox{-90}{$\rightarrow$}}

         \put(-130,138){\rotatebox{-135}{$\rightarrow$}}
         \put(-105,134){\rotatebox{-45}{$\rightarrow$}}

         \put(-115,60){\small \rotatebox{90}{$\text{wormhole}$}}

         \put(-465,50){\small \rotatebox{90}{$\text{left boundary}$}}
         \put(-225,50){\small \rotatebox{90}{$\text{left boundary}$}}

         \put(-270,120){\small \rotatebox{-90}{$\text{right boundary}$}}
         \put(-5,120){\small \rotatebox{-90}{$\text{right boundary}$}}

\end{tabular}
\end{center}
\caption{ \small Schematic representation of the $t=0$ slice of (a) the unperturbed two-sided black brane geometry and (b) the two-sided black brane geometry in the presence of a shock wave. We assume that the shock wave is sent at some time $t_0<0$, therefore, it effectively increases the size of the wormhole at $t=0$. In both cases the blue curves represent the U-shaped extremal surfaces $\gamma_A$ (in the left side of the geometry) and $\gamma_B$ (in the right side of the geometry). The red curves represent extremal surfaces $\gamma_1$ and $\gamma_2$ connecting the two sides of the geometry. The extremal surface $\gamma_\text{wormhole}$ defined in the text is given by the union of these two surfaces, $\gamma_\text{wormhole}=\gamma_1 \cup \gamma_2$.}
\label{fig-surfaces}
\end{figure}

In order to compute the two-sided mutual information we consider a strip-like region $A$ on the left boundary of the geometry and an identical region $B$ on the right boundary. The mutual information is defined as
\be
I(A,B)=S_A+S_B-S_{A \cup B}\,,
\ee
where $S_A$ is the entanglement entropy of region $A$, and so on. The above entanglement entropies can be computed holographically using the HRRT prescription (\ref{eq-RT}). The first two terms, $S_A$ and $S_B$, are given by the area of the U-shaped extremal surfaces $\gamma_{A,B}$ whose boundary coincide with the boundary of $A$ and $B$, respectively. These surfaces lie outside the event horizon, in the left and right regions, respectively. The last term, $S_{A \cup B}$, is given by the area of the extremal surface whose boundary coincides with the boundary of $A \cup B$. There are two candidates for this extremal surface. The first one is the surface $\gamma_A \cup \gamma_B$, while the second one is a surface $\gamma_\text{wormhole}$ that stretches through the wormhole connecting the two boundaries of the geometry. See figure \ref{fig-surfaces} for a schematic illustration. If the surface $\gamma_A \cup \gamma_B$ has less area than the surface $\gamma_\text{wormhole}$, then we have $I(A,B)=0$, because $\text{Area}(\gamma_A \cup \gamma_B)=\text{Area}(\gamma_A)+\text{Area}(\gamma_B)$. On the other hand, if $\gamma_\text{wormhole}$ has less area than $\gamma_A \cup \gamma_B$, then we have that $\text{Area}(\gamma_A \cup \gamma_B)<\text{Area}(\gamma_A)+\text{Area}(\gamma_B)$, which implies a positive mutual information $I(A,B)>0$.

Before proceeding further, let us explain the general expectations. In the unperturbed geometry, the mutual information must be zero if the regions $A$ and $B$ are small enough, and become positive for large regions. The presence of the shock wave should decrease the amount of mutual information at a given time slice $t>t_0$. Eventually, the mutual information must drop to zero as we move the shock wave farther into the past $t_0\to-\infty$. As explained in the introduction, the positive mutual information characterizes the special left-right pattern of entanglement of the TFD state, and the fact that $I(A,B)$ decreases (and eventually vanishes) in a shock wave geometry shows that this pattern of entanglement is sensitive to arbitrarily small perturbations sent in the asymptotic past.

\subsection{Mutual information in the TFD state}

We will discuss two cases, the ``commutative strip'' and the ``non-commutative strip'', as defined in reference \cite{Fischler-2013}.

\subsubsection*{Commutative strip:}
The region dubbed as the ``commutative strip'' are the set of points with $0 \leq x_1 \leq \ell$ and $-L/2 \leq x_{2,3} \leq L/2$, with $L\to\infty$.
In this case the appropriate embedding is $X^m =(0,x(r),x_2,x_3,r,\theta_i)$, where $\theta_i$ are the angles of the five-sphere. The components of the induced metric are the following:
\bea
&g_{22}& = g_{33} = G_{22}\,,\\
&g_{\theta_i \theta_i}&=G_{\theta \theta} \times \text{metric on $S^5$}\,,\\
&g_{rr}& = G_{rr}+G_{11}\,x'(r)^2\,.
\eea
The area functional to be extremized is given by
\bea
\text{Area}(\gamma_A)&=&\int d^8 \sigma \sqrt{\text{det}\,g_{ab}}\,,\\
&=&\Omega_5  \int dx_2\,dx_3\,dr\,G_{22}\,G_{\theta \theta}^{5/2}\,\left(G_{rr}+G_{11}\,x'(r)^2 \right)^{1/2}\,,\\
&=&\Omega_5 L^2 R^8 \int dr r^3 \left( \frac{1}{r^4 f}+x'(r)^2 \right)^{1/2}\,,\\
&=& \Omega_5 L^2 R^8 \int dr\, \mathcal{L}(x,x';r)\,,
\eea
where $L^2 = \int dx_2\,dx_3$, and $\Omega_5$ is the volume of a unit $S^5$. The above functional does not depend on $x$, and so there is a conserved quantity associated to translations in $x$
\be
p= \frac{\partial \mathcal{L}}{\partial x'}=\frac{r^3 x'}{\sqrt{\frac{1}{r^4 f}+x'^2}} = r_m^3\,,
\label{eq-gammaC}
\ee
where in the last equality we computed $p$ at the point $r=r_m$ at which $x' \rightarrow \infty$.
By solving the equation (\ref{eq-gammaC}) for $x'$ we get
\be
x'^2=\frac{1}{r^4 f} \frac{1}{\left(\frac{r^6}{r_m^6}-1 \right)}\,.
\label{eq-x1prime}
\ee
Using equation (\ref{eq-x1prime}) we can write the on-shell area of the surface as
\be
\text{Area}(\gamma_A) = 2 \Omega_5 L^2 R^8 \int_{r_m}^{\infty} dr\,\frac{r}{\sqrt{f}}\frac{1}{\sqrt{1-r_m^6/r^6}}\,,
\ee
so, the entanglement entropies of the subregions $A$ and $B$ are
\be
S_A = S_B=\frac{\text{Area}(\gamma_A)}{4 G_\mt{N}}= \frac{\Omega_5 L^2 R^8}{2\,G_\mt{N}}  \int_{r_m}^{\infty} dr\,\frac{r}{\sqrt{f}}\frac{1}{\sqrt{1-r_m^6/r^6}}\,.
\ee
The entanglement entropy $S_{A \cup B}$ is computed from the area of the extremal surface $\gamma_\text{wormhole}$ connecting the two sides of the geometry
\be
\text{Area}(\gamma_\text{wormhole}) = 4 \Omega_5 L^2 R^8 \int_{\rh}^{\infty} dr\,\frac{r}{\sqrt{f}}\,,
\ee
where the factor of 4 comes from the fact that we have two sides in the geometry and two disconnected surfaces, at $x_1=0$ and $x_1=\ell$, respectively. The entanglement entropy of $A \cup B$ is then given by
\be
S_{A \cup B}=\frac{\text{Area}(\gamma_\text{wormhole})}{4 G_\mt{N}}=\frac{\Omega_5 L^2 R^8}{G_\mt{N}} \int_{\rh}^{\infty} dr\,\frac{r}{\sqrt{f}}\,.
\ee
From the above expressions we can compute the mutual information,
\be
I(A,B)=\frac{\Omega_5 L^2 R^8}{G_\mt{N}} \left[\int_{r_m}^{\infty} dr\,\frac{r}{\sqrt{f}}\frac{1}{\sqrt{1-r_m^6/r^6}}-\int_{\rh}^{\infty} dr\,\frac{r}{\sqrt{f}} \right]\,,
\ee
which is a function of the turning point $r_m$. We can plot the mutual information as a function of the strip's width $\ell$ by writing the later quantity as a function of $r_m$,
\be
\ell = \int dx = \int x' dr = 2 \int_{r_m}^{\infty} \frac{dr}{r^2}\frac{1}{\sqrt{f}\sqrt{r^6/r_m^6-1}}\,,
\ee
and then making a parametric plot of $I(A,B)$ versus $\ell$. Note, however, that both quantities $I(A,B)$ and $\ell$ are independent of the non-commutative parameter ``$a$''. This means that the results for the commutative strip are the same as for a strip in a 5-dimensional AdS-Schwarzschild geometry. The plot of $I(A,B)$ for the commutative strip is shown in figure \ref{fig-MIversusL}, and corresponds to the curve labeled by $a=0$ (black curve).

\subsubsection*{Non-commutative strip:}
\begin{figure}[t!]
\begin{center}
\setlength{\unitlength}{1cm}
\includegraphics[width=0.6\linewidth]{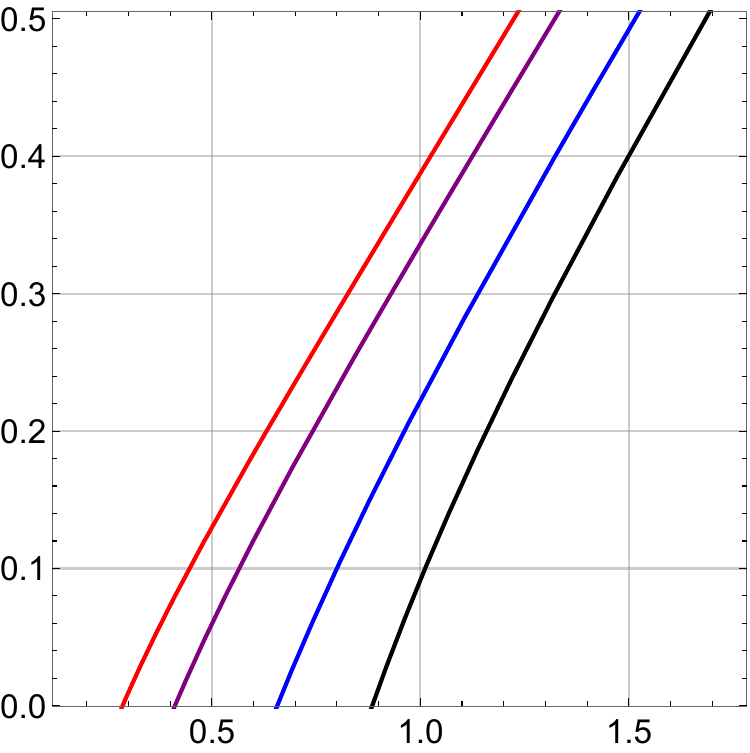}
\put(-4,-.5){\large $\ell$}
\put(-9.7,+4.1){\rotatebox{90}{\large $I(A,B)$}}

\end{center}
\caption{ \small
Mutual Information (in units of $L^2 R^3/G_\mt{N}^{(5)}$) as a function of the strip's width $\ell$ for non-commutative SYM theory. The curves correspond from the right to the left to $a = 0$ (black curve), $a = 0.8$ (blue curve), $a = 1.2$ (purple curve) and  $a = 1.2$ (red curve) . In all cases we have fixed $\rh=1$.}
\label{fig-MIversusL}
\end{figure}

The ``non-commutative strip'' is given by the set of points with $0 \leq x_2 \leq \ell$ and $-L/2 \leq x_{1,3} \leq L/2$, with $L\to\infty$.
In this case the appropriate embedding is $X^m =(0,x_1,x(r),x_3,r,\theta_i)$ and the components of the induced metric are
\bea
&g_{11}&= G_{11}\,,\\
&g_{33}&= G_{22}\,,\\
&g_{\theta_i \theta_i}&=G_{\theta \theta} \times \text{metric on $S^5$}\,,\\
&g_{rr}& = G_{rr}+G_{22}\,x'(r)^2\,.
\eea
The area functional to be extremized is given by
\bea
\text{Area}(\gamma_A)&=&\int d^8 \sigma \sqrt{\text{det}\,g_{ab}}\,,\\
&=&\Omega_5  \int dx_1\,dx_3\,dr\,G_{11}^{1/2}\,G_{22}^{1/2}\,G_{\theta \theta}^{5/2}\,\left(G_{rr}+G_{22}\,x'(r)^2 \right)^{1/2}\,,\\
&=&\Omega_5 L^2 R^8 \int dr r^3 \left( \frac{1}{r^4 h f}+x'(r)^2 \right)^{1/2}\,,\\
&=& \Omega_5 L^2 R^8 \int dr\, \mathcal{L}(x,x';r)\,,
\eea
where $L^2 = \int dx_1\,dx_3$, and $\Omega_5$ is the volume of a unit $S^5$. The above functional does not depend on $x$, and so there is a conserved quantity associated to translations in $x$
\be
p= \frac{\partial \mathcal{L}}{\partial x'}=\frac{r^3 x'}{\sqrt{\frac{1}{r^4 h f}+x'^2}} = r_m^3\,,
\label{eq-gammaC}
\ee
where in the last equality we computed $p$ at the point $r=r_m$ at which $x' \rightarrow \infty$.
By solving the equation (\ref{eq-gammaC}) for $x'$ we get
\be
x'^2=\frac{1}{r^4 h f} \frac{1}{\left(\frac{r^6}{r_m^6}-1 \right)}\,.
\label{eq-x2prime}
\ee
Using equation (\ref{eq-x2prime}) we can write the on-shell area of the surface as
\be
\text{Area}(\gamma_A) = 2 \Omega_5 L^2 R^8 \int_{r_m}^{\infty} dr\,\frac{r}{\sqrt{h f}}\frac{1}{\sqrt{1-r_m^6/r^6}}\,,
\ee
so, the entanglement entropies of the subregions $A$ and $B$ are
\be
S_A = S_B=\frac{\text{Area}(\gamma_A)}{4 G_\mt{N}}= \frac{\Omega_5 L^2 R^8}{2\,G_\mt{N}}  \int_{r_m}^{\infty} dr\,\frac{r}{\sqrt{h f}}\frac{1}{\sqrt{1-r_m^6/r^6}}\,.
\ee
The entanglement entropy $S_{A \cup B}$ is computed from the area of the extremal surface $\gamma_\text{wormhole}$ connecting the two sides of the geometry
\be
\text{Area}(\gamma_\text{wormhole}) = 4 \Omega_5 L^2 R^8 \int_{\rh}^{\infty} dr\,\frac{r}{\sqrt{h f}}\,,
\ee
where the factor of 4 comes from the fact that we have two sides in the geometry and two disconnected surfaces, at $x_2=0$ and $x_2=\ell$, respectively. The entanglement entropy of $A \cup B$ is then given by
\be
S_{A \cup B}=\frac{\text{Area}(\gamma_\text{wormhole})}{4 G_\mt{N}}=\frac{\Omega_5 L^2 R^8}{G_\mt{N}} \int_{\rh}^{\infty} dr\,\frac{r}{\sqrt{h f}}\,.
\ee
From the above expressions we can compute the mutual information,
\be
I(A,B)=\frac{\Omega_5 L^2 R^8}{G_\mt{N}} \left[\int_{r_m}^{\infty} dr\,\frac{r}{\sqrt{h f}}\frac{1}{\sqrt{1-r_m^6/r^6}}-\int_{\rh}^{\infty} dr\,\frac{r}{\sqrt{h f}} \right]\,,
\ee
which is a function of the turning point $r_m$. We can plot the mutual information as a function of the strip's width $\ell$ by writing the later quantity as a function of $r_m$,
\be
\ell = \int dx = \int x' dr = 2 \int_{r_m}^{\infty} \frac{dr}{r^2}\frac{1}{\sqrt{h f}\sqrt{r^6/r_m^6-1}}\,,
\ee
and then making a parametric plot of $I(A,B)$ versus $\ell$. Both quantities $I(A,B)$ and $\ell$ depend on the non-commutative parameter ``$a$'', because they have factors of $h(r)=(1+a^4r^4)^{-1}$. Also, note that we can recover the expressions for the commutative strip by setting $a=0$ (or equivalently $h=1$). In figure \ref{fig-MIversusL} we plot $I(A,B)$ as a function of the strip's width $\ell$ for several values of the non-commutative parameter at a fixed temperature. As expected from the results of mutual information for the one-sided black brane \cite{Fischler-2013}, increasing in the non-commutative parameter $a$ reduces the critical length $\ell=\ell_{\text{crit}}$ and hence lowers the threshold for the phase transition of mutual information. This implies that non-commutativity introduces more correlations between two sub-systems as compared to the commutative case.

\subsection{Disruption of mutual information by shock waves} \label{sec-disruptMI}
Let us now study how the two-sided mutual information changes in the presence of a shock wave geometry. In the following, we will specialize to the case of a homogeneous shock wave, in which the shock wave parameter has the form $\alpha=\text{constant}\,\times e^{2\pi t_0/\beta}$. The mutual information in a shock wave geometry will be denoted as
\be
I(A,B;\alpha)=S_A+S_B-S_{A \cup B}(\alpha)\,,
\ee
where we have indicated that $S_{A \cup B}$ generically depends on the shock wave parameter $\alpha$, while $S_A$ and $S_B$ do not. This can be easily understood, since
the corresponding extremal surfaces $\gamma_A$ and $\gamma_B$ remain outside the horizon, while the shock wave only affects quantities that probe the black hole interior.

As expected on general grounds, the entanglement entropy $S_{A \cup B}(\alpha)$ has various $\alpha$-independent divergences. In practice we find it convenient to define a regularized entanglement entropy
\be
S_{A \cup B}^\text{reg}(\alpha)=S_{A \cup B}(\alpha)-S_{A \cup B}(\alpha=0)\,,
\ee
and rewrite the mutual information as
\be
I(A,B;\alpha)=S_A+S_B-S_{A \cup B}(\alpha)=I(A,B;\alpha=0)-S_{A \cup B}^\text{reg}(\alpha)\,.
\ee

In the following, we will consider cases where $\ell>\ell_{\text{crit}}$ so that the mutual information is positive in the unperturbed geometry, i.e., $I(A,B;\alpha=0)>0$. That means that the extremal surface stretching between the two sides of the geometry $\gamma_\text{wormhole}$ has smaller area than the two extremal surfaces lying outside the black brane $\gamma_A$ and $\gamma_B$. When $\alpha > 0$, the wormhole becomes longer and the area of the extremal surface probing the interior also increases, resulting in a decrease of the mutual information. As $\alpha$ increases the mutual information eventually drops to zero, signaling the total disruption of two-sided correlations.

\subsubsection*{Commutative strip:}
\begin{figure}[t!]
\centering

\begin{tikzpicture}[scale=1.5]

\draw [thick,decorate,decoration={zigzag,segment length=2mm, amplitude=0.5mm}]  (5.5,3) -- (9.5,3);

\draw [thick,decorate,decoration={zigzag,segment length=2mm, amplitude=0.5mm}]  (6.6,0) -- (10.6,0);

\draw [thick] (6.5,0) -- (9.5,3);
\draw [thick] (6.6,0) -- (9.6,3);

\draw [thick] (5.5,3) -- (6.5,0);

\draw [thick] (9.6,3) -- (10.6,0);

\draw [thick,dashed] (5.5,3) -- (7.5,1);
\draw [thick,dashed] (8.6,2) -- (10.6,0);

\draw[very thick, red] (6,1.5) -- (10.1,1.5);

\draw[thick,red, |-|] (7,1.5) -- (7.5,1.5);

\draw[thick,red, |-|] (7,1.5) -- (8,1.5);

\draw [thick,<->] (8.13,1.4) -- (7.63,.9);

\node [scale=.8] at (8.,1.0) {$\frac{\alpha}{2}$};

\draw [dashed, blue] (5.5,3) to (7.1,1.65) to [out=-35,in=180] (7.5,1.52)
to [out=0,in=-145] (7.9,1.65) to (9.5,3);

\node [scale=.7] at (7.5,1.7) {$r_0$};

\node [scale=.45] at (6.5,1.39) {$I$};
\node [scale=.45] at (7.27,1.39) {$II$};
\node [scale=.45] at (7.73,1.39) {$III$};

\end{tikzpicture}
\vspace{0.1cm}
\caption{ \small Extremal surface (horizontal, red) in the shock wave geometry. We divide the left half of the surface into three parts, $I$, $II$ and $III$. The segments $II$ and $III$ have the same area and they are separated by the point $r_0$ at which the constant-$r$ surface (blue, dashed curve, defined by $r=r_0$) intersects the extremal surface.}
\label{fig-surfaceLocation}
\end{figure}
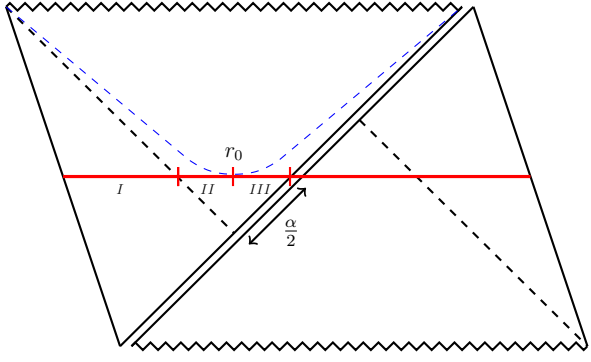

The appropriate embedding in this case is $X^m=(t,0,x_2,x_3,r(t),\theta_i)$. The components of the induced metric are
\bea
&g_{22}&=g_{33}=G_{22}\,,\\
&g_{\theta_i \theta_i}&=G_{\theta \theta} \times \text{metric on $S^5$}\,,\\
&g_{tt}&=G_{tt}+G_{rr}\dot{r}^2\,,
\eea
and the functional to be extremized is
\bea
\text{Area}(\gamma_\text{wormhole})&=&2\Omega_5 \int dt dx_2 dx_3 G_{22}\,G_{\theta \theta}^{5/2}\left(G_{tt}+G_{rr}\dot{r}^2 \right)^{1/2}\,,\\
&=& 2\Omega_5 L^2 R^8 \int dt\,r^3 \left(-f+\frac{\dot{r}^2}{f r^4} \right)^{1/2}\,,\\
&=& 2\Omega_5 L^2 R^8 \int dt\,\mathcal{L}(r,\dot{r};t)\,.
\eea
Since the above functional is invariant under $t$-translations, there is an associated conserved quantity,
\be
\mathcal{E}=\frac{\partial \mathcal{L}}{\partial \dot{r}}\dot{r}-\mathcal{L}=\frac{r^3 f}{\sqrt{-f+\frac{\dot{r}^2}{f r^4}}}=-r_0^3 \sqrt{-f(r_0)}\,,
\label{eq-gammaCsw}
\ee
where in the last equality we computed $\mathcal{E}$ at the point $r_0$ at which $\dot{r}=0$. By solving (\ref{eq-gammaCsw}) for $\dot{r}$ we obtain
\be
\dot{r}^2 = \left( r^2 f \right)^2 \left( 1+\mathcal{E}^{-2} f r^{6}\right)\,.
\label{eq-rdotC}
\ee
Using the above result we can write the on-shell area as
\be
\text{Area}(\gamma_\text{wormhole})= 2\Omega_5 L^2 R^8 \int dr  \frac{r^2}{\sqrt{\mathcal{E}^2 r^{-4}+r^2 f}}\,,
\ee
and the time coordinate $t$ along the extremal surface as
\be
t(r)= \int dt = \int \frac{dr}{\dot{r}}=\int \frac{dr}{r^2 f \sqrt{1+\mathcal{E}^{-2}f r^6}}\,.
\ee
Since these expressions do not depend on the non-comutative parameter $a$, we can expect the disruption of mutual information
to be the same as for the commutative SYM theory. However, we will proceed with the analysis for illustrative purposes.
The entanglement entropy $S_{A \cup B}$ is given by
\be
S_{A \cup B} = \frac{2\Omega_5 L^2 R^8}{4 G_\mt{N}} \int dr \frac{r^2}{\sqrt{\mathcal{E}^2 r^{-4}+r^2 f}}\,.
\ee
It is convenient to divide the region of integration of the above integral into three regions, $I$, $II$ and $III$, as shown in figure \ref{fig-surfaceLocation}. Since the regions $II$ and $III$ have the same area, we can write $\int_{I \cup II \cup III}=\int_{\rh}^{\infty}+2\int_{r_0}^{\rh}$. The entanglement entropy $S_{A \cup B}$ can then be written more explicitly as
\be
S_{A \cup B}(r_0) = \frac{\Omega_5 L^2 R^8}{G_\mt{N}} \left[ \int_{\rh}^{\infty} dr \frac{r^2}{\sqrt{\mathcal{E}^2 r^{-4}+r^2 f}}+2\int_{r_0}^{\rh} dr \frac{r^2}{\sqrt{\mathcal{E}^2 r^{-4}+r^2 f}} \right],
\ee
where the extra factor of 2 accounts for the two sides of the geometry.
The effect of the shock wave on $S_{A \cup B}$ is controlled by the turning point $r_0\leq\rh$. The shock wave is absent when $r_0 = \rh$ (or, equivalently, $\mathcal{E}=0$), and its effects become stronger as one decreases $r_0$. In terms of this parameter, the regularized entanglement entropy $S_{A \cup B}^\mt{reg}$ can be written as
\bea
S_{A \cup B}^\mt{reg}(r_0)&=&S_{A \cup B}(\alpha)-S_{A \cup B}(\alpha=0)=S_{A \cup B}(r_0)-S_{A \cup B}(\rh)\\
&=&\frac{\Omega_5 L^2 R^8}{G_\mt{N}} \left[ \int_{\rh}^{\infty} dr \left( \frac{r^2}{\sqrt{\mathcal{E}^2 r^{-4}+r^2 f}} -\frac{r^2}{\sqrt{r^2 f}}\right)+2\int_{r_0}^{\rh} dr \frac{r^2}{\sqrt{\mathcal{E}^2 r^{-4}+r^2 f}} \right].\nonumber
\eea
Finally, the shock wave parameter $\alpha$ can also be written as a function of the turning point $r_0$ (see appendix \ref{appA} for details). The final result reads
\be
\alpha(r_0)=2\, e^{K_1(r_0)+K_2(r_0)+K_3(r_0)}\,,
\ee
where
\bea
&K_1&= \frac{4 \pi}{\beta}  \int_{\bar{r}}^{r_0} dr \frac{1}{r^2 f}\,,\\
&K_2&= \frac{2 \pi}{\beta}  \int_{\rh}^{\infty} dr \frac{1}{r^2 f} \left(1-\frac{1}{\sqrt{1+ \mathcal{E}^{-2}f r^6}} \right)\,,\\
&K_3&= \frac{4 \pi}{\beta} \int_{r_0}^{\rh} dr \frac{1}{r^2 f} \left(1-\frac{1}{\sqrt{1+\mathcal{E}^{-2} f r^6} } \right)\,.
\eea
As expected, the shock wave parameter $\alpha = \alpha(r_0)$ increases monotonically as one decreases $r_0$ (with $\alpha(\rh)=0$) and it diverges at some critical radius $r_c=\frac{\rh}{3^{1/4}}$ (see appendix \ref{appB} for details). Indeed, both $K_3$ and $t(r_0)\,\propto \,t_0$ diverge at $r_0=r_c$. This means that the region $r<r_c$ and, in particular, the singularity cannot be probed by $I(A,B;\alpha)$, even in the limit $t_0\to\infty$. The results for the commutative strip can be found by setting $a=0$ in the results for the non-commutative strip (which will be presented below). The plots for $\alpha(r_0)$, as well as for $S_{A \cup B}^\mt{reg}(\alpha)$ and $I(A,B,\alpha)$ for the commutative strip are shown in figures \ref{fig-alpha} and \ref{fig-spreading}, respectively, and correspond to the curves labeled by $a=0$ (black curves).

\subsubsection*{Non-commutative strip:}
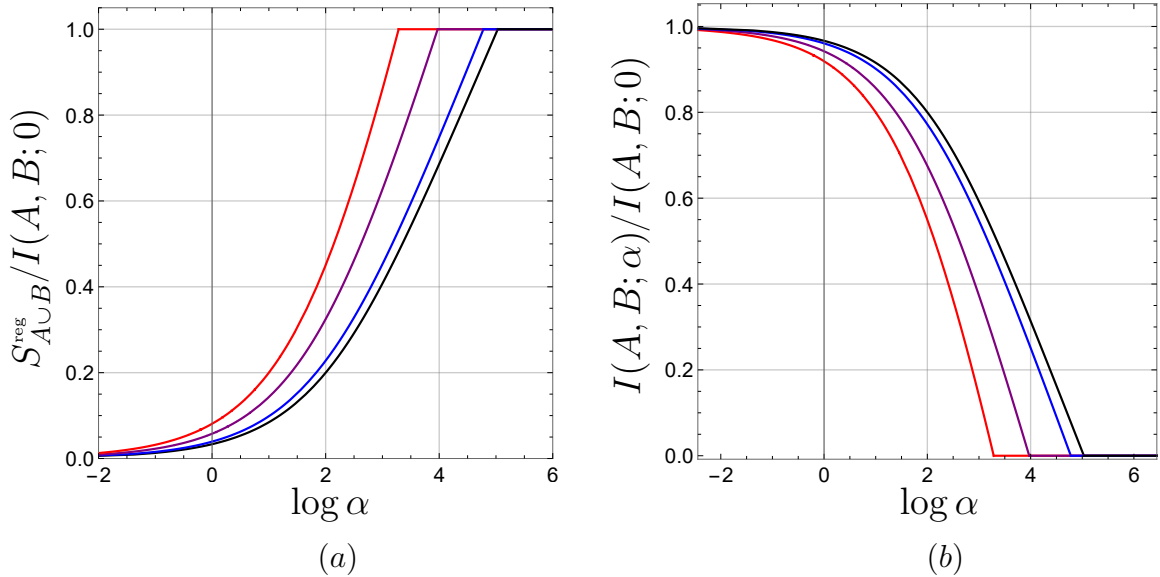
\begin{figure}[t!]
\begin{center}
\begin{tabular}{cc}
\setlength{\unitlength}{1cm}
\hspace{-0.9cm}
\includegraphics[width=6.5cm]{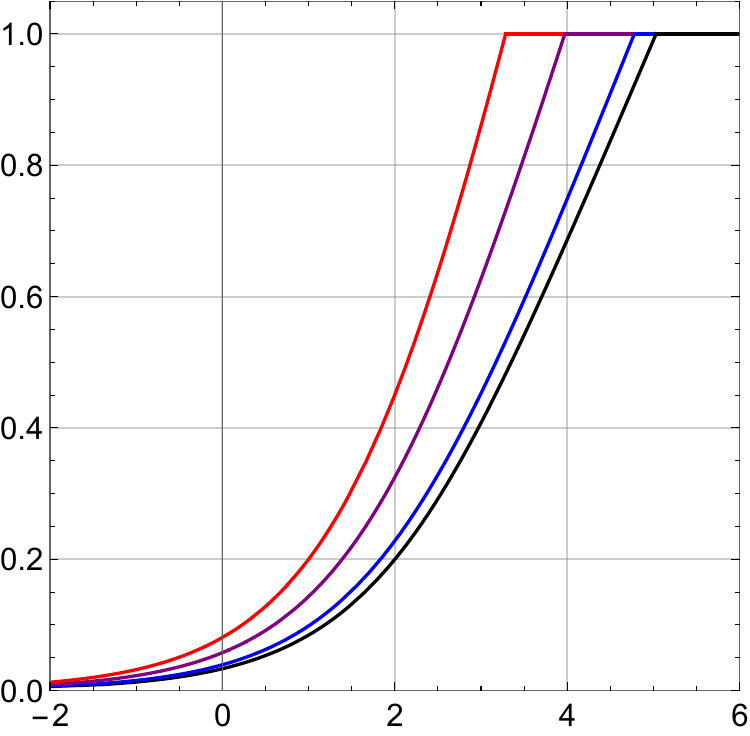}
\qquad\qquad &
\includegraphics[width=6.5cm]{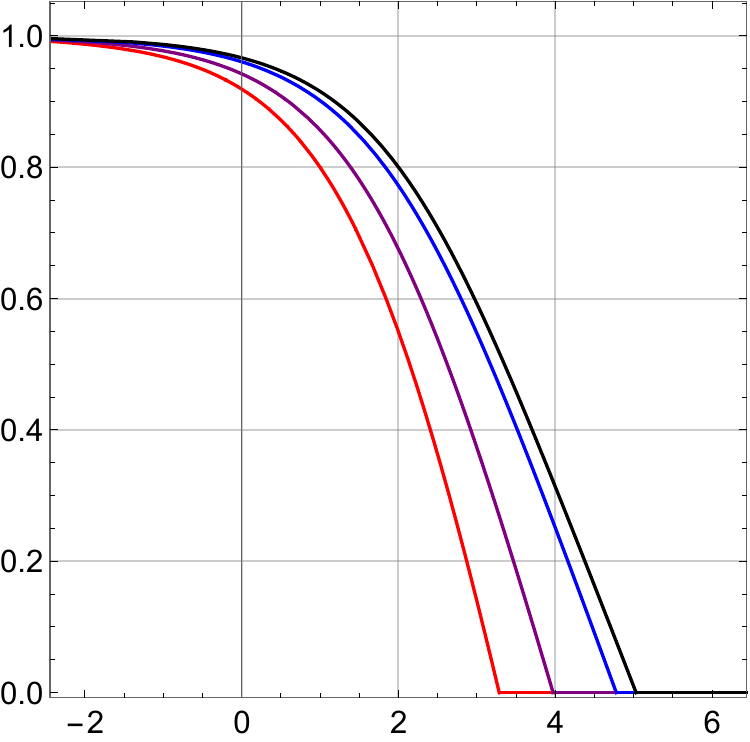}
\qquad
  \put(-430,45){\rotatebox{90}{\large $S_{A \cup B}^\mt{reg}/I(A,B;0)$}}
         \put(-325,-10){\large $\log \alpha$}
         \put(-205,35){\rotatebox{90}{\large $I(A,B;\alpha)/I(A,B;0)$}}
         \put(-97,-10){\large $\log \alpha$}
         \put(-87,-30){$(b)$}
         \put(-315,-30){$(a)$}

\end{tabular}
\end{center}
\caption{ \small (a) Regularized entanglement entropy $S_{A \cup B}^\mt{reg}$ and (b) mutual information $I(A,B)$ as a function of $ \log \alpha$. Both in (a) and (b) the curves correspond to $a= 0$ (black curves), $a= 0.8$ (blue curves), $a=1.2$ (purple curves) and $a = 1.5$ (red curves). In all cases we have fixed $\rh=1$.}
\label{fig-spreading}
\end{figure}

The appropriate embedding in this case is $X^m=(t,x_1,0,x_3,r(t),\theta_i)$. The components of the induced metric are
\bea
&g_{11}&=G_{11}\,,\\
&g_{33}&=G_{22}\,,\\
&g_{\theta_i \theta_i}&=G_{\theta \theta} \times \text{metric on $S^5$}\,,\\
&g_{tt}&=G_{tt}+G_{rr}\dot{r}^2\,,
\eea
and the functional to be extremized is
\bea
\text{Area}(\gamma_\text{wormhole})&=&2\Omega_5 \int dt\, dx_1 \,dx_3\, G_{11}^{1/2 }\,G_{22}^{1/2}\,G_{\theta \theta}^{5/2}\left(G_{tt}+G_{rr}\dot{r}^2 \right)^{1/2}\,,\\
&=& 2\Omega_5 L^2 R^8 \int dt\,\frac{r^3}{h^{1/2}} \left(-f+\frac{\dot{r}^2}{f r^4} \right)^{1/2}\,,\\
&=& 2\Omega_5 L^2 R^8 \int dt\,\mathcal{L}(r,\dot{r};t)\,.
\eea
Since the above functional is invariant under $t$-translations, there is an associated
conserved quantity,
\be
\mathcal{E}=\frac{\partial \mathcal{L}}{\partial \dot{r}}\dot{r}-\mathcal{L}=\frac{r^3 f h^{-1/2}}{\sqrt{-f+\frac{\dot{r}^2}{f r^4}}}=-r_0^3 \sqrt{-\frac{f(r_0)}{h(r_0)}}\,,
\label{eq-gammaNCsw}
\ee
where in the last equality we computed $\mathcal{E}$ at the point $r_0$ at which $\dot{r}=0$. By solving (\ref{eq-gammaNCsw}) for $\dot{r}$ we obtain
\be
\dot{r}^2 = \left( r^2 f \right)^2 \left( 1+\mathcal{E}^{-2}  r^{6}\frac{f}{h}\right)\,.
\label{eq-rdotC}
\ee
Using the above result we can write the on-shell area as
\be
\text{Area}(\gamma_\text{wormhole})= 2\Omega_5 L^2 R^8 \int dr \frac{r^2}{h^{1/2}\sqrt{\mathcal{E}^2 r^{-4}h+r^2 f}}\,,
\ee
and the time coordinate $t$ along the extremal surface as
\be
t(r)= \int dt = \int \frac{dr}{\dot{r}}=\int \frac{dr}{r^2 f \sqrt{1+\mathcal{E}^{-2}r^6 \frac{f}{h}}}\,.
\ee
The entanglement entropy $S_{A \cup B}$ is then given by
\be
S_{A \cup B} = \frac{2\Omega_5 L^2 R^8}{4 G_\mt{N}} \int dr \frac{r^2}{h^{1/2}\sqrt{\mathcal{E}^2 r^{-4}h+r^2 f}}\,.
\ee
Again, we divide the region of integration of the above integral into three regions, $I$, $II$ and $III$, as shown in figure \ref{fig-surfaceLocation}.
Since the regions $II$ and $III$ have the same area, we can write $\int_{I \cup II \cup III}=\int_{\rh}^{\infty}+2\int_{r_0}^{\rh}$. The entanglement entropy $S_{A \cup B}$ can then be written more explicitly as
\be
S_{A \cup B}(r_0) = \frac{\Omega_5 L^2 R^8}{G_\mt{N}} \left[ \int_{\rh}^{\infty} dr\frac{r^2}{h^{1/2}\sqrt{\mathcal{E}^2 h r^{-4}+r^2 f}}+2\int_{r_0}^{\rh} dr \frac{r^2}{h^{1/2}\sqrt{\mathcal{E}^2 h r^{-4}+r^2 f}} \right]\,.
\ee
where the extra factor of 2 accounts for the two sides of the geometry. The effect of the shock wave on $S_{A \cup B}$ is controlled by the turning point $r_0\leq\rh$. The shock wave is absent when $r_0 = \rh$ (or, equivalently, $\mathcal{E}=0$), and its effects become stronger as one decreases $r_0$. In terms of this parameter, the regularized entanglement entropy $S_{A \cup B}^\mt{reg}$ can be written as
\bea
S_{A \cup B}^\mt{reg}(r_0)&=&S_{A \cup B}(\alpha)-S_{A \cup B}(\alpha=0)=S_{A \cup B}(r_0)-S_{A \cup B}(\rh)\,,\\
&=&\frac{\Omega_5 L^2 R^8}{G_\mt{N}} \bigg[ \int_{\rh}^{\infty} dr \left( \frac{r^2}{h^{1/2}\sqrt{\mathcal{E}^2 h r^{-4}+r^2 f}} -\frac{r^2}{h^{1/2}\sqrt{r^2 f}}\right)\nonumber\\
&&\qquad\qquad\qquad\qquad+2\int_{r_0}^{\rh} dr\frac{r^2}{h^{1/2}\sqrt{\mathcal{E}^2 h r^{-4}+r^2 f}} \bigg]. \nonumber
\eea
Finally, the shock wave parameter $\alpha$ can be written as a function of the turning point $r_0$ (see appendix \ref{appA} for details). The final result reads
\be
\alpha(r_0)=2\, e^{K_1(r_0)+K_2(r_0)+K_3(r_0)}\,,
\ee
where
\bea
&K_1&= \frac{4 \pi}{\beta}  \int_{\bar{r}}^{r_0} dr \frac{1}{r^2 f}\,,\label{K1NC}\\
&K_2&= \frac{2 \pi}{\beta}  \int_{\rh}^{\infty} dr \frac{1}{r^2 f} \left(1-\frac{1}{\sqrt{1+ \mathcal{E}^{-2}f h^{-1} r^6}} \right)\,,\\
&K_3&= \frac{4 \pi}{\beta} \int_{r_0}^{\rh} dr \frac{1}{r^2 f} \left(1-\frac{1}{\sqrt{1+\mathcal{E}^{-2} f h^{-1} r^6} } \right)\,.\label{K3NC}
\eea
\begin{figure}[t!]
\begin{center}
\setlength{\unitlength}{1cm}
\includegraphics[width=0.6\linewidth]{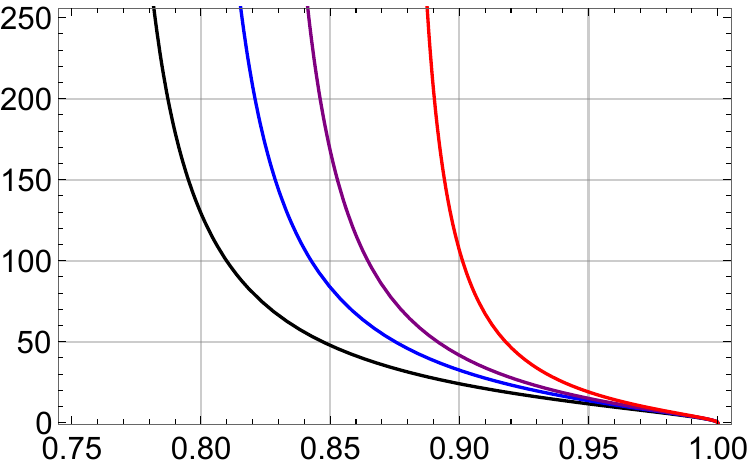}
\put(-4.5,-.5){\large $r_0/\rh$}
\put(-10.3,+3.2){\large $\alpha(r_0)$}

\end{center}
\caption{ \small
Shock wave parameter $\alpha$ versus the `turning point' $r_0$ divided by $\rh$ for non-commutative SYM theory. The curves correspond to $a = 0$ (black curve), $a = 0.8$ (blue curve), $a = 1$ (purple curve) and $a = 2$ (red curve). In all cases we have fixed $\rh=1$.}
\label{fig-alpha}
\end{figure}
The shock wave parameter $\alpha = \alpha(r_0)$ increases monotonically as one decreases $r_0$ (with $\alpha(\rh)=0$) and it diverges at some critical radius
\be
r_c = \frac{\rh}{10^{1/4}} \left(-\frac{3}{a^4\rh^4}+3+\frac{\sqrt{9+2 a^4 \rh^4+9 a^8 \rh^8}}{a^4\rh^4} \right)^{1/4}\,.
\label{eq-rcritical}
\ee
Indeed both $K_3$ and $t(r_0)\,\propto\,t_0$ diverge at $r_0=r_c$. In figure \ref{fig-alpha} we plot the shock wave parameter $\alpha$ versus the `turning point' $r_0$ for
several values of the non-commutative parameter. In general, we observe that $r_0$ gets repealed from the singularity as we increase the strength of the non-commutativity, meaning that the extremal surface probe less of the interior. This might be a consequence of the fuzzy nature of the non-commutative geometry. In figure \ref{fig-spreading} we show the behavior of the regularized entanglement entropy $S_{A \cup B}^\mt{reg}(\alpha)$ in a shock wave geometry and how this results in the disruption of the two-sided mutual information $I(A,B;\alpha)$. As we can see from these plots, the disruption of the mutual information occurs faster as we increase the non-commutative parameter $a$. In the next section we will quantify this statement more clearly by the calculation of the so-called entanglement velocity.

\subsubsection{Entanglement velocity}\label{sec:VE}

Before saturation, the entanglement entropy $S_{A \cup B}(\alpha)$ grows linearly with $\log \alpha$, and this implies that it grows linearly with the time $t_0$ at which the system was perturbed ($\alpha = \text{const}\times e^{\frac{2\pi}{\beta}t_0}$). From this linear behaviour we can define the so-called {\it entanglement velocity}, which is a quantity that characterizes the spread of entanglement in chaotic system. In the following, we will specialize to the case of the non-commutative strip. The results for the commutative strip can be obtained simply by setting $a=0$ in all the formulas below.

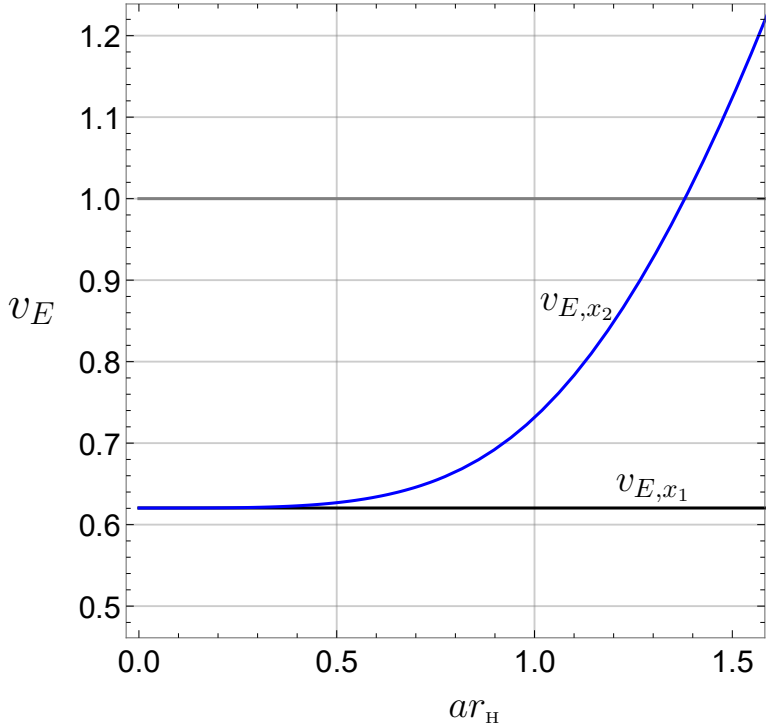
\begin{figure}[t!]
\begin{center}
\setlength{\unitlength}{1cm}
\includegraphics[width=0.6\linewidth]{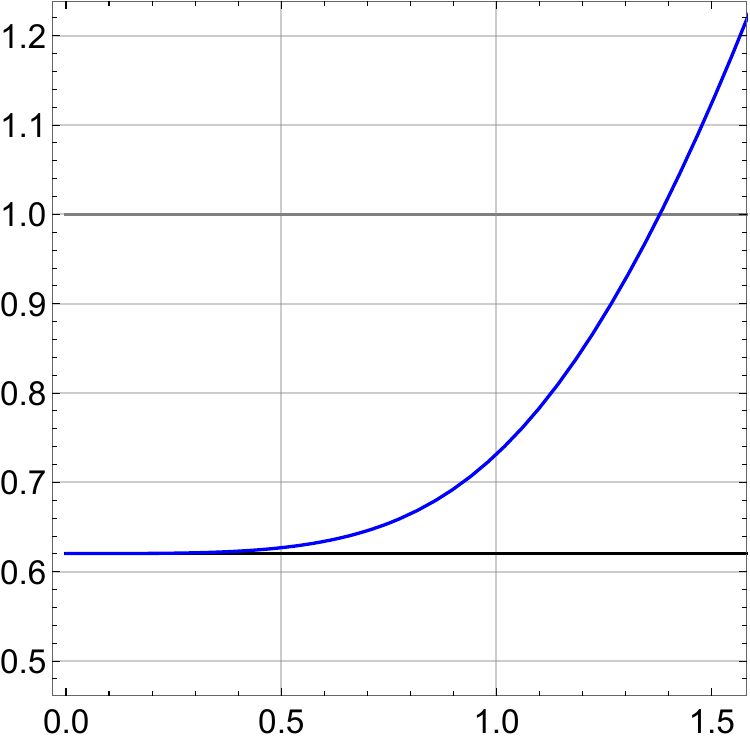}
\put(-4.2,-.5){\large $a \rh$}
\put(-10.,+4.8){\Large $v_E$}
\put(-3.,+4.9){\large $v_{E,x_2}$}
\put(-2.,+2.5){\large $v_{E,x_1}$}
\end{center}
\caption{ \small
Entanglement velocity $v_E$ versus the dimensionless parameter $a \rh$. The blue curve represent the entanglement velocity for a non-commutative strip, while the horizontal black line represent the entanglement velocity for a commutative strip, which is equal to the conformal result $v_E=\sqrt{2}/3^{3/4}$. The horizontal grey line represents the speed of light.}
\label{fig-VE}
\end{figure}

As shown in the previous section, the function $\alpha(r_0)$ increases monotonically as we decrease $r_0\leq\rh$ and diverges at a critical radius $r_0=r_c$ given by (\ref{eq-rcritical}).  In the vicinity of $r_c$, one can show that
\be
S_{A \cup B}^\mt{reg} \cong \frac{2\Omega_5 L^2 R^8}{G_\mt{N}} r_c^3 \sqrt{-\frac{f(r_c)}{h(r_c)}} \,\frac{\beta}{4 \pi}\, \text{log}\,\alpha\,, \,\,\,\, \text{for}\,\,\,\, r_0 \approx r_c\,.
\label{Slog}
\ee
Since the shift $\alpha$ grows exponentially with time, $\alpha = \text{constant} \times e^{2\pi t_0/\beta}$, the above result implies that $S_{ A \cup B}$ grows linearly with $t_0$. The rate of change of $S_{A \cup B}^\mt{reg}$ with the shock wave time is
\be
\frac{d S_{A \cup B}^\mt{reg}}{dt_0}  = \frac{L^2 R^3}{G_\mt{N}^{(5)}} \frac{r_c^3 \sqrt{-f(r_c)}}{\sqrt{h(r_c)}}\,,
\ee
where $G_\mt{N}^{(5)}=\frac{G_\mt{N}}{\Omega_5 R^5}$ is the five-dimensional Newton constant. Using the formula for the thermal entropy density,
\be
s_\mt{th}=\frac{R^3 \rh^3}{4 G_\mt{N}^{(5)}}\,,
\ee
we can rewrite the above equation as
\be
\frac{d S_{A \cup B}^\mt{reg}}{dt_0} = s_\mt{th} A_\Sigma  \left( \frac{r_c^3}{\rh^3}\sqrt{\frac{-f(r_c)}{h(r_c)}} \right),
\ee
where $A_\Sigma=4L^2$ is the area of the 4 hyperplanes defining $\Sigma=\partial (A\cup B)$. Finally, comparing with the formula (\ref{vEdef}) we can then extract the entanglement velocity for the non-commutative strip, which we denote as
\be
v_{E,x_2} =  \frac{r_c^3}{\rh^3}\sqrt{\frac{-f(r_c)}{h(r_c)}}\,.
\label{eq-VEperp}
\ee
One can check that by setting $a=0$ in this formula we obtain the standard entanglement velocity for a strip in ordinary SYM theory,
\be
v_{E,x_1}=v_{E,x_2}(a=0)=\frac{\sqrt{2}}{3^{3/4}}\,,
\ee
which also applies for the commutative strip in non-commutative SYM. More generally, expanding in powers of $a \rh$ we obtain,
\be
v_{E,x_2}=\frac{r_c^3}{\rh^3}\sqrt{\frac{-f(r_c)}{h(r_c)}}=\frac{\sqrt{2}}{3^{3/4}}+\frac{a^4 \rh^4}{3 \sqrt{2} 3^{3/4}}+\frac{5 a^8 \rh^8}{108 \sqrt{2} 3^{3/4}}+\mathcal{O}(a^{12} \rh^{12})\,,
\ee
which shows that $v_{E,x_2}$ increases as we increase the non-comutative parameter! In figure \ref{fig-VE} we plot both $v_{E,x_1}$ and $v_{E,x_2}$ as a function of $a \rh$. We observe that $v_{E,x_2}$ exceeds the speed of light already at some value of $a\rh$ of order $a\rh\sim\mathcal{O}(1)$ and grows without bound in the limit of strong non-commutativity. This behavior is in qualitative agreement with the results obtained for the entanglement entropy for a free scalar field on the fuzzy sphere following a quantum quench \cite{Sabella-Garnier:2017svs}. Finally, we note that $v_{E,x_i}\leq v_{B,x_i}$ generically, for any value of the non-commutative parameter. This implies that the conjecture made in \cite{Mezei-2016} holds for our non-commutative setup, and suggests that it might indeed be true for any (possibly non-local) quantum system.

\section{Butterfly velocity from one-sided perturbations} \label{sec-VB}
In this section we present an alternative derivation of $v_{B}$ that does not rely on the shock wave results. This alternative way of computing $v_{B}$ is based on entanglement wedge subregion duality \cite{Headrick-2014}, and it was first proposed in \cite{Mezei-2016}. Here we extend their results for the kind of anisotropic metrics that we consider in this paper.

\subsection{Infalling particle and entanglement wedge}

The derivation goes as follows. Consider the application of a localized bulk operator $V$ in a black brane geometry. This operator creates a one particle state in the bulk theory which eventually falls into the black hole and thermalizes. As the particle falls into the black hole, $V$ gets scrambled with an increasing number of degrees of freedom and, as a result, the operator effectively grows in space. This is consistent with the standard intuition from the holographic UV/IR connection \cite{Susskind:1998dq,Peet:1998wn,Agon:2014rda}, which implies that the information of the particle gets delocalized over a larger region as it falls deeper into the bulk. The proposal of \cite{Mezei-2016} is that, at late times, the rate of growth of this region is controlled by the butterfly velocity.

In this context, the butterfly velocity can be calculated using the entanglement wedge subregion duality. According to this duality a certain subregion $A$ of the boundary theory can be completely described by a subregion in the bulk geometry, which is called the entanglement wedge of $A$. In the following, we compute the butterfly velocity by requiring that the entanglement wedge of a certain region $A$ contains the particle created by $V$.

Let us assume a generic black hole metric of the form
\be
ds^2 = - G_{tt}dt^2+G_{rr}dr^2+G_{ij}dx^i dx^j\,,\,\
\ee
where $i,j=1,2,...,D$. Here $(t,x^i)$ with  $i=1,...,d-1$ are the boundary theory coordinates, and $r$ is the holographic radial coordinate. The coordinates $x^i$ with $i=d, d+1,...,D$ describe a compact space, which can affect $v_\mt{B}$ if $G_{ij}$ depend on $r$ for $i,j=d, d+1,...,D$. We note that this is indeed true in the non-commutative setup that we consider, in which case the compact space is given by $G_{\theta\theta}(r)d\Omega^2_5$. We assume the boundary is located at $r=\infty$ and the horizon at $r=\rh$. We now consider a fixed time slice of the geometry, at a long time after the application of $V$, such that this operator is delocalized in a very large region $A$. This limit simplifies the analysis for two reasons. First, the equations of motion defining the entanglement wedge linearise, because the corresponding RT surface lies very close to the black hole horizon. Second, the particle created by $V$ also lies very close to the horizon, having a simple description in terms of Rindler coordinates.

We assume the following near-horizon expressions
\be
G_{tt} = c_0 (r-\rh)\,,\,\,\,\,G_{rr}=\frac{c_1}{r-\rh}\,,\,\,\,\,G_{ij} =G_{ij}(\rh)+G_{ij}'(\rh)(r-\rh)\,.
\ee
In terms of $c_0$ and $c_1$, the inverse Hawking temperature reads
\be
\beta=4 \pi \sqrt{\frac{c_1}{c_0}}\,.
\ee
It is convenient to go to Rindler coordinates, $\rho^2=(r-\rh)\left( \frac{2 \pi}{\beta} \right)^2\frac{1}{G_{tt}'(\rh)}$, in which terms the above metric becomes
\be
ds^2 = -\left( \frac{2 \pi}{\beta} \right)^2 \rho^2 dt^2 +d\rho^2+\left[G_{ij}(\rh)+\frac{G_{ij}'(\rh)}{G_{tt}'(\rh)}\left( \frac{2 \pi}{\beta} \right)^2 \rho^2 \right]dx^i dx^j\,.
\ee
The infalling particle gets blue shifted as it falls into the black hole and, at late times, it approaches the horizon exponentially
\be
\rho(t)=\rho_0 e^{-\frac{2\pi}{\beta}t}\,.
\ee
Now we proceed to calculate the position of the RT surface defining the entanglement wedge of $A$.
Considering the embedding $X^{m}=(0,x^i,\rho(x^i))$, the area functional can be written as
\be
\text{Area} = \sqrt{\det{G_{ij}(\rh)}} \int d^{d-1}x \left[1+\left( \frac{2 \pi}{\beta} \right)^2 \frac{\rho^2}{G_{tt}'(\rh)} G^{ii}(\rh) G_{ii}'(\rh)+G^{ii}(\rh)(\partial_i \rho)^2 \right],
\ee
where we have assumed $G_{ij}$ to be diagonal. The equations of motion that follows from the above functional are
\be
G^{ii}(\rh)\partial_i^2 \rho(x^i) = M^2 \rho(x^i)\,,
\ee
where
\be
M^2= \left( \frac{2 \pi}{\beta} \right)^2 \frac{G^{ii}(\rh)G_{ii}'(\rh)}{G_{tt}'(\rh)}=  \left( \frac{2\pi}{\beta} \right)^2 \frac{1}{G_{tt}'(\rh)} \left[ \frac{G_{11}'}{G_{11}}+2\frac{G_{22}'}{G_{22}}+5 \frac{G_{\theta \theta}'}{G_{\theta \theta}}\right]\,,
\ee
In order to solve this equation we define the new coordinates $\sigma^i =x^i/\sqrt{G^{ii}(\rh)}$, in which terms the equation of motion becomes
\be
\left( \frac{\partial}{\partial \sigma^i} \right)^2 \rho = M^2 \rho\,.
\ee
The solution of the above equation is \cite{Mezei-2016}
\be
\rho(\sigma^i)=\rho_\text{min} \frac{\Gamma(a+1)}{2^{-a}M^{a}} \frac{I_a(M |\sigma|)}{|\sigma|^a}\qquad\text{with}\qquad a=(d-3)/2\,.
\ee
In this formula, $\rho_\text{min}$ is interpreted as the radius of closest approach to the horizon and $I_a$ is a modified Bessel function of the second kind. As explained in \cite{Mezei-2016}, when $\rho$ exceeds $\beta$, the surface exits the near horizon region and reaches the boundary very quickly. It is then possible to determine the size of the region $A$ in terms of $\rho_\text{min}$ by solving the equation
\be
\beta=\rho_\text{min} \frac{\Gamma(a+1)}{2^{-2}\mu^{a}} \frac{I_a(M R_{\sigma})}{|R_{\sigma}|^a}\,,
\ee
where $R_{\sigma}$ is the size of the region $A$ in the $\sigma$-coordinates. The approximate solution for this equation at large $R_{\sigma}$ is
\be
\rho_\text{min} \approx e^{-M R_{\sigma}}\,.
\ee

\subsubsection{Butterfly velocity\label{sec:vB2}}

For an anisotropic system, $v_\mt{B}$ is different along the different directions and so is the size of the region $A$. Let us say that $R_i$ is the size of the region $A$ along the $x^i$ direction. This is related to its size in $\sigma$-coordinates by the equation $R_{\sigma i}=\sqrt{G_{ii}(\rh)}R_i$. Requiring the infalling particle created by $V$ to be contained in the entanglement wedge implies
\be
\rho_\text{min} \leq \rho(t)\qquad \Rightarrow\qquad M \sqrt{G_{ii}(\rh)}R_i \geq \frac{2\pi}{\beta}t\,,
\ee
or, equivalently
\be
R_i \geq v_{B,x_i} t\,,
\ee
where the butterfly velocity $v_{B,x_i}$ along the $x^i$-direction is calculated as
\be
v_{B,x_i}=\frac{2\pi}{\beta}\frac{1}{\sqrt{G_{ii}(\rh)} M}=\frac{\sqrt{G_{tt}'(\rh)}}{\sqrt{G_{ii}(\rh)}\sqrt{G^{kk}(\rh)G_{kk}'(\rh)}}\,.
\ee
This formula is in complete agreement with the ones obtained from shock wave calculations (\ref{VBpara})-(\ref{VBperp}).

Before closing this section we would like to offer some intuition about the boundary picture of the bulk operator $V$ and the corresponding one particle state, in our non-commutative setup. Without loss of generality, we can imagine inserting $V$ directly at the boundary, and follow the evolution of the created particle as it falls into the black brane. In ordinary AdS/CFT this would mean that we are turning on an operator localized in space, which can in turn be interpreted as a local quench, see e.g. \cite{Nozaki:2013wia}. However, in non-commutative gauge theories there are no local operators in position space, so it is necessary to explain how the above prescription works in the present case.

As explained in section \ref{sec-gaugeinv}, there is a natural set of gauge invariant operators that can be defined in non-commutative gauge theories, which can be obtained by smearing the ordinary gauge covariant operators $\mathcal{O}(x)$ over a Wilson line $W$, according to (\ref{wline}). The size of this Wilson line $\ell_W$ scales with the momentum $k$, roughly as $\ell_W \simeq \theta k $. Let us now imagine having a very large Wilson line along one of the directions, say the $x_1$-direction. This can be achieved by taking either $\theta$ or $k_1$ to be very large, such that $\ell_W \rightarrow \infty$. In this approximation the information about the perturbation is initially localized along the $x_1$-axis. As the system evolves in time, the information will get delocalized in a cylindrical region around this axis. In the bulk, the information about this perturbation will be contained inside the entanglement wedge, which will also display a cylindrical symmetry.
The derivation of $v_B$ goes as before, except that now the entanglement surface will not depend on $x_1$. In particular, the butterfly velocity will be the same as before and will describe how fast the information about the smeared operator gets delocalized inside a `cylinder' whose radius along the $\vec{x}$-direction is $v_{B}(\phi)\, t$, where $\phi$ is the angle between $\vec x$ and the $x_1$-direction. In more general cases, when the Wilson line is not very large, we expect $v_B(\phi)$ to describe the expansion of the operator in a region around the Wilson line that defines it.

\section{Conclusions and outlook} \label{sec-discussion}

In this paper we have studied shock waves in the gravity dual to $\mathcal{N}=4$ non-commutative SYM theory.  From the shock wave profiles, we extracted several chaos-related properties of this system, namely, the butterfly velocity, the scrambling time, and the Lyapunov exponent. As expected on general grounds, we find that the Lyapunov exponent saturates the chaos bound, $\lambda_L = 2\pi/\beta$, while the scrambling time scales logarithmically with the entropy of the system, $t_* = \frac{\beta}{2\pi} \log S$. Since neither the temperature nor the entropy are affected by the non-commutativity, both $\lambda_L$ and $t_*$ are exactly the same as the corresponding values in ordinary SYM theory.

In contrast, the butterfly velocity is largely affected by the non-commutative parameter $\theta$, specially in the UV.
The results for $v_B$ as a function of $a\,\rh=\pi\lambda^{1/4}\sqrt{\theta}T$ are shown in figure \ref{fig-VB2}. Since the non-commutativity
is introduced along the $x_2-x_3$ plane, i.e. $[x_2,x_3]\sim i\theta$, the gravity dual is hence anisotropic, $G_{11} \neq G_{22}=G_{33}$. This causes $v_B$ to depend on the direction of the perturbation. For simplicity, we only computed the components $v_{B,x_1}$ and $v_{B,x_2}$, where the first one is the butterfly velocity along the $x_1$-direction, and the latter one corresponds to the butterfly velocity along the $x_2$- and $x_3$- directions. We observe that, at some $a \,\rh \sim \mathcal{O}(1)$,  $v_{B,x_2}$ becomes larger than the speed of light, while $v_{B,x_1}$ remains subluminal. In fact $v_{B,x_1}$ takes the universal conformal value for all $a \,\rh $.

The fact that along the non-commutative directions the butterfly velocity exceeds the speed of light in the regime of strong non-locality is not surprising. Indeed, Lorentz invariance is explicitly broken by the non-commutative parameter $\theta$ so the standard notions of causality do not apply. Nevertheless, this result is remarkable in the context of quantum information theory, since it represents a novel violation of the known bounds on the rate of transfer of information. We comment, though, that in this limit the information is highly delocalized due to the UV/IR mixing,\footnote{Recall that in non-commutative theories, the information of any degree of freedom moving with a large momentum is highly delocalized in the transverse directions $\ell_\perp\sim \theta k$ \cite{Bigatti:1999iz}.} so the implementation of a local protocol to retrieve the information might require an exponentially longer time than the commutative case. As a result, an increase on $v_B$ due to the non-commutativity is necessarily compensated by an increase in the ``computational cost'' or a decrease on the amount of ``useful information'' at fixed  time. It would be interesting to understand this phenomenon better.

Finally, we also computed the entanglement velocity $v_E$ by studying the disruption of the two-sided mutual information in the presence of homogeneous shock waves. In figure \ref{fig-spreading} we show the results for $S_{A\cup B}$ and $I(A,B)$ for various values of the non-commutative parameter. In general, we find that the mutual information is reduced in the presence of the shock wave, and eventually vanishes as one let $t_0\to \infty$. Right before the transition, the entanglement entropy of the two sub-systems $S_{A\cup B}$ grows linearly, with a slope given by $v_E$. In figure \ref{fig-VE} we show the behavior of the entanglement velocity as a function of the non-commutative parameter. We considered two geometries. For a ``commutative strip'' (strip with finite width along the $x_1$-direction), the results are the same as for an AdS black brane, while for a ``non-commutative'' strip (strip with finite width along the $x_2$- or $x_3$-direction) the entanglement velocity increases with the non-commutative parameter. Eventually, $v_{E,x_2}$ exceeds the speed of light in the limit of strong non-locality. This behavior is in qualitative agreement with the results obtained for the entanglement entropy for a free scalar field on the fuzzy sphere following a quantum quench \cite{Sabella-Garnier:2017svs}.

We also confirmed the expectation based on the conjecture proposed in \cite{Mezei-2016}, namely that $v_{E,x_i}\leq v_{B,x_i}$ in general quantum systems. Indeed, we find that this is valid in our setup for any value of the non-commutative parameter suggesting that the conjecture might indeed be true for any (possibly non-local) quantum system. It would be interesting to test further this conjecture in other non-local theories, for example, in the gravity dual of the dipole deformation of $\mathcal{N}=4$ super Yang Mills \cite{Bergman:2001rw} or in the gravity  dual of the so-called little string theory \cite{Aharony:1998ub}. It will also interesting to compute $v_B$ directly in a non-commutative field theory (using perturbation theory) and compare with the strong coupling results obtained in this paper.

\acknowledgments

It is a pleasure to thank Jan de Boer, Jose Edelstein, Philippe Sabella-Garnier and Koenraad Schalm for useful discussions and comments on the manuscript.
The research of WF is based upon work supported by the National Science
Foundation under Grant Number PHY-1620610. VJ is supported by Mexico's National
Council of Science and Technology (CONACyT) under grant CB-2014/238734. JFP is supported by the Netherlands Organization for Scientific Research (NWO) under the VENI grant 680-47-456/1486.

\appendix

\section{Momentum space correlator and the butterfly velocity} \label{appA0}
In this appendix we study shock wave geometries and present a detailed derivation of the formula (\ref{eq-VBani}). We start from the solution of the shock wave profile $\tilde{\alpha}$ at finite momentum,
\be
\tilde{\alpha}(t,\vec k)= \frac{e^{2 \pi (t-t_*) / \beta}}{ G^{ii}k_i^2+M^2}\,.
\ee
The above function has a pole at $ G^{ii}k_i^2+M^2=0$. By writing the momentum components in spherical coordinates
\bea
k_1&=&\sqrt{G_{11}(\rh)}k \cos \phi\,,\\
k_2&=&\sqrt{G_{22}(\rh)}k \sin \phi \sin \phi_2\,,\\
k_3&=&\sqrt{G_{33}(\rh)}k \sin \phi \cos \phi_2\,,
\eea
the position of the pole can be specified as $k=i M$. Interestingly, at the pole, the modulus of $\vec k$ gives us the ratio of the Lyapunov exponent and the butterfly velocity
\bea
k^2&=&k_1^2+k_2^2+k_3^2=-\lambda_L^2\, \mu^2 \left(G_{11} \cos^2\phi+G_{22} \sin^2 \phi \sin^2 \phi_2+G_{33} \sin^2 \phi \cos^2 \phi_2 \right)\,,\nonumber \\
&=&-\frac{\lambda_L^2}{v_{B}^2(\phi,\phi_2)}\,,
\eea
where $\lambda_L=2\pi/\beta$ is the Lyapunov exponent and
\be
v_B(\phi,\phi_2)=\frac{1}{\mu\,\sqrt{G_{11} \cos^2\phi+G_{22} \sin^2 \phi \sin^2 \phi_2+G_{33} \sin^2 \phi \cos^2 \phi_2} }\,,
\ee
is the butterfly velocity along an arbitrary direction. Note that, in the most general case, $G_{11}\neq G_{22} \neq G_{33}$, the butterfly velocity is completely anisotropic and depends on the two spherical angles $\phi$ and $\phi_2$. If we assume isotropy in the $x_2-x_3$ plane, i.e. $G_{33}=G_{22}$, the above formula simplifies to
\be
v_B(\phi)=\frac{1}{\mu\,\sqrt{G_{11} \cos^2\phi+G_{22} \sin^2 \phi} }\,,
\ee
which leads to (\ref{eq-VBani}). Note that $v_B$ still depends on $\phi$. This is a consequence of the anisotropy in the $x_1$-direction, i.e. $G_{11} \neq G_{22}$.

The fact that the pole of $C(t,\vec{k})$ gives the Lyapunov exponent and the butterfly velocity is implicit in other holographic calculations (see for instance the Appendix C of \cite{Gu:2016oyy}). In our setup, we can confirm that the quantity appearing in the pole of $\tilde{\alpha}(t,\vec k)$ can indeed be identified with the butterfly velocity. We do so by looking at the limit $k\to0$ (or $|\vec x|>\!\!>\sqrt{\theta}$), in which the size of the Wilson line is vanishingly small $\ell_W\to0$ and the shock wave becomes approximately local. In this limit we can write the shock wave profile in position space as the Fourier transform of $\tilde{\alpha}$,
\be
\alpha(t,\vec x)=\int \frac{d^3\vec{k}}{(2\pi)^3} \frac{e^{\frac{2 \pi}{\beta} (t-t_*)}e^{i \vec{k} \cdot \vec{x}}}{ G^{ii}k_i^2+M^2}\,.
\label{alpha-k}
\ee
By changing variables as $k_i \rightarrow \sqrt{G_{ii}}q_i$, the above integral can be written as\footnote{In the last equality we use that $\int \frac{d^3q}{(2\pi)^3}e^{i \vec{\sigma} \cdot \vec{q}} \frac{4\pi}{q^2+M^2}=\frac{e^{-M|\vec \sigma|}}{|\vec \sigma|}$.}
\be
\alpha(t,\vec \sigma)=\sqrt{G_{11}G_{22}G_{33}}\int \frac{d^3\vec{q}}{(2\pi)^3} \frac{e^{\frac{2 \pi}{\beta} (t-t_*)}e^{i \vec{q} \cdot \vec{\sigma}}}{|\vec{q}\,|^2+M^2}=\frac{\sqrt{G_{11}G_{22}G_{33}}}{4\pi}\frac{e^{\frac{2 \pi}{\beta} (t-t_*)}e^{-M |\vec \sigma|}}{|\vec \sigma|}\,,
\label{sol-alpha-sigma}
\ee
where all the metric functions are evaluated at the horizon. To further simplify this expression, we write
\be
M^2=\left( \frac{2\pi}{\beta} \right)^2 \frac{G^{ii}(\rh)G_{ii}'(\rh)}{G_{tt}'(\rh)} \equiv \left( \frac{2\pi}{\beta} \right)^2 \mu^2\,.
\ee
With the above definitions we can write
\be
M |\vec{\sigma}|=\frac{2\pi}{\beta} \sqrt{\sum_i \mu^2 (x^i)^2 G_{ii} } \equiv \frac{2\pi}{\beta} \frac{|\vec{x}|}{v_B(\phi,\phi_2)}\,,
\label{eq-M}
\ee
where the angles $(\phi,\phi_2)$ are defined such that
\bea
x_1&=&|\vec x| \cos \phi\,,\\
x_2&=&|\vec x| \sin \phi \sin \phi_2\,,\\
x_3&=&|\vec x| \sin \phi \cos \phi_2\,.
\eea
Substituting (\ref{eq-M}) in (\ref{sol-alpha-sigma}) we obtain
\be
\alpha(t,\vec x) = e^{\frac{2\pi}{\beta}\left(t-t_*-\frac{|\vec x|}{v_B(\phi,\phi_2)} \right)}\,,
\ee
which is the well known shock wave profile for the case of localized perturbations. This confirms that that the quantity
$v_B(\phi,\phi_2)$ appearing at the pole of $\tilde{\alpha}(t,\vec k)$ it is indeed the butterfly velocity.

\section{Shock wave parameter $\alpha$ as a function of $r_0$} \label{appA}

In the case of homogeneous shocks, the strength of the shock wave can be either measured by the parameter $\alpha$ or by the `turning point' $r_0$. In this appendix determine the relation between these two parameters. In the following we will specialize to the case of the non-commutative strip. The case of a commutative strip can be recovered from our results by setting $a = 0$ in all the formulas below.

By symmetry considerations we know that the extremal surface whose area gives $S_{A \cup B}$ divides the bulk into two halves, as shown in figure \ref{fig-surfaceLocation}. The parameter $r_0$ defines a constant-$r$ surface inside the black horizon which intersect the extremal surface exactly at the point at which $\dot{r}=0$. We split the left part of the surface into three segments $I$, $II$ and $III$. The first segment goes from the boundary $(U,V)=(1,-1)$ to the horizon $(U,V)=(U_1,0)$. The second segment goes from the horizon $(U,V)=(U_1,0)$ to the point $(U,V)=(U_2,V_2)$ where the extremal surface intersects with the constant-$r$ surface at $r=r_0$. The third segment connects the point $(U,V)=(U_2,V_2)$ to the horizon at $(U,V)=(0,\alpha/2)$. In what follows we compute the unknown quantities $U_1, U_2$ and $V_2$ in terms of $r_0$ and obtain an expression for $\alpha(r_0)$.

In the left exterior region, the Kruskal coordinates are defined as
\be
U=e^{\frac{2\pi}{\beta}(r_*-t)}\,, \,\,\,\, V=-e^{\frac{2\pi}{\beta}(r_*+t)}\,,\,\,\,\,r_{*}=-\int_{r}^{\infty} dr' \frac{1}{r'^2 f(r')}\,,
\ee
while, inside the black hole and in the right side of the geometry, these coordinates are defined as
\be
U=e^{\frac{2\pi}{\beta}(r_*-t)}\,,\,\,\,\,  V=e^{\frac{2\pi}{\beta}(r_*+t)}\,,\,\,\,\, r_{*}=\int_{\bar{r}}^{r} dr' \frac{1}{r'^2 f(r')}\,,
\ee
where $\bar{r}$ is a point behind the horizon at which $r_* = 0$. On the other hand, the time $t(r)$ along the extremal surface can be written as
\be
t(r)= \int \frac{dr}{r^2 f \sqrt{1+\mathcal{E}^{-2}r^6 h^{-1}f}}\,.
\ee
Using the above equations we can express variation in the coordinates $U$ and $V$ as
\bea
&\Delta \log U^2 &= \frac{4\pi}{\beta}\left(\Delta r_*-\Delta t \right)=\frac{4\pi}{\beta} \int dr\, \frac{1}{r^2 f}\left(\frac{1}{\sqrt{1+\mathcal{E}^{-2}r^6 h^{-1}f}}-1 \right) \,,\cr
&\Delta \log V^2 &= \frac{4\pi}{\beta}\left(\Delta r_*+\Delta t \right)=\frac{4\pi}{\beta} \int dr\, \frac{1}{r^2 f}\left(\frac{1}{\sqrt{1+\mathcal{E}^{-2}r^6 h^{-1}f}}+1\right) \,.
\eea
The coordinate $U_1$ can be calculated considering the variation of $U$ from the boundary to the horizon
\be
U_1^2 = \text{exp}\left[\frac{4\pi}{\beta} \int_{\rh}^{\infty} dr\, \frac{1}{r^2 f}\left(\frac{1}{\sqrt{1+\mathcal{E}^{-2}r^6 h^{-1}f}}-1 \right) \right]\,.
\ee
To compute $U_2$ we consider the variation of $U$ from $r=\rh$ to $r=r_0$
\be
\frac{U_2^2}{U_1^1} = \text{exp}\left[ \frac{4\pi}{\beta} \int_{r_0}^{\rh} du\,\frac{1}{r^2 f}\left(\frac{1}{\sqrt{1+\mathcal{E}^{-2}r^6 h^{-1}f}}-1 \right) \right]\,.
\ee
The coordinate $V_2$ can be written as
\be
V_2=\frac{1}{U_2} \, \text{exp}\left[ \frac{4\pi}{\beta} \int_{\bar{r}}^{r_0} dr\,\frac{1}{r^2 f}\right]\,.
\ee
The shift $\alpha$ can then be computed by considering the variation in the $V$-coordinate along the segment $III$
\be
\frac{\alpha^2}{4 V_2^2}=\text{exp}\left[\frac{4\pi}{\beta} \int_{\rh}^{r_0} du\,\frac{1}{r^2 f}\left(\frac{1}{\sqrt{1+\mathcal{E}^{-2}r^6 h^{-1}f}}-1\right) \right]=\frac{U_1^2}{U_2^2}\,.
\ee
Finally, after some simplifications we find that the parameter $\alpha$ can be expressed as
\be
\alpha(r_0)=2\, e^{K_1(r_0)+K_2(r_0)+K_3(r_0)}\,,
\ee
where the $K_i$'s are given by equations (\ref{K1NC})-(\ref{K3NC}).

\section{Divergence of $K_3(r_0)$} \label{appB}

In this appendix we determine the critical radius $r_0 =r_c$ at which $K_3(r_0)$ diverges. According to (\ref{K3NC}), $K_3(r_0)$ is given by
\be
K_3= \frac{4 \pi}{\beta} \int_{r_0}^{\rh} dr \frac{1}{r^2 f} \left(1-\frac{1}{\sqrt{1+\mathcal{E}^{-2} f h^{-1} r^6} } \right)\,.
\label{eq-K3NC}
\ee
The critical radius $r_c$ can be obtained by considering the integrand of the above equation in the limit $r \rightarrow r_0$. Notice that in this limit
\bea
\mathcal{E}^{-2} f h^{-1} r^6 &=&-\frac{f h^{-1} r^6}{f(r_0) h(r_0)^{-1} r_0^6} \nonumber\\
& = & \frac{f(r_0) h(r_0)^{-1} r_0^6+\big( f h^{-1} r^6 \big)'\big|_{r=r_0} (r-r_0)}{f(r_0) h(r_0)^{-1} r_0^6}+\mathcal{O}(r-r_0)^2\nonumber \\
&=& 1+\frac{\big( f h^{-1} r^6\big)'}{ f h^{-1} r^6}\Big|_{r=r_0}(r-r_0)+\mathcal{O}(r-r_0)^2\,.
\eea
Using the above result in equation (\ref{eq-K3NC}) one finds
\be
K_{3}\approx -\frac{4 \pi}{\beta} \int_{r_0}^{\rh} dr \frac{1}{r_0^2f(r_0)} \left(1-\frac{1}{\sqrt{-\frac{\big( f h^{-1} r^6 \big)'}{ f h^{-1} r^6}\Big|_{r=r_0}(r-r_0)}} \right)\,.
\ee
Indeed, the above expression diverges when $r_0 \rightarrow r_c$ such that
\be
\frac{\big(f h^{-1} r^6\big)'}{ f h^{-1} r^6}\Big|_{r=r_c}=0\,.
\ee
The solution to this equation is given by equation (\ref{eq-rcritical}) and approaches the standard value $r_c\to\frac{\rh}{3^{1/4}}$ in the limit $a\to0$.

\end{document}